\documentclass[a4paper]{article}

\usepackage{jheppub}
\usepackage{slashed}
\usepackage{subfig}
\usepackage{xcolor}
\usepackage{booktabs}
\usepackage{adjustbox}
\usepackage{ulem}

\makeatletter
\gdef\@fpheader{}
\makeatother

\newcommand{\Zp}{{Z^\prime}}

\newcommand{\gp}{{g^\prime}}
\newcommand{\gl}{{g_\ell}}
\newcommand{\gt}{{g_t}}

\newcommand{\SM}{{\rm SM}}
\newcommand{\NP}{{\rm NP}}

\newcommand{\GeV}{{\,\rm GeV}}
\newcommand{\TeV}{{\,\rm TeV}}

\newcommand{\fb}{{\, \rm fb}}

\newcommand{\mB}{{\mathcal B}}
\newcommand{\mO}{{\mathcal O}}
\newcommand{\mC}{{\mathcal C}}

\newcommand{\Phil}{{\Phi_\ell}}
\newcommand{\Phit}{{\Phi_t}}

\newcommand{\nn}{\nonumber}

\allowdisplaybreaks[4]

\title{\bf \boldmath Correlating the CDF $W$-mass shift with the muon $g-2$ and the $b \to s \ell^+ \ell^-$ transitions}

\author[a,b]{Xin-Qiang Li,}
\emailAdd{xqli@mail.ccnu.edu.cn}

\author[a]{Ze-Jun Xie,}
\emailAdd{xiezejun@mails.ccnu.edu.cn}

\author[a,c]{Ya-Dong Yang}
\emailAdd{yangyd@ccnu.edu.cn}

\author[a]{and Xing-Bo Yuan}
\emailAdd{y@ccnu.edu.cn}

\affiliation[a]{Institute of Particle Physics and Key Laboratory of Quark and Lepton Physics~(MOE),\\
Central China Normal University, Wuhan, Hubei 430079, China}
\affiliation[b]{Center for High Energy Physics, Peking University, Beijing 100871, China}
\affiliation[c]{Institute of Particle and Nuclear Physics, Henan Normal University, Xinxiang 453007, China}

\abstract{\noindent
Motivated by the latest CDF $W$-mass measurement as well as the muon $g-2$ anomaly and the discrepancies observed in $b \to s \ell^+ \ell^-$ transitions, we propose an extension of the Standard Model (SM) with the $SU(2)_L$-singlet vector-like fermion partners that are featured by additional $U(1)^\prime$ gauge symmetry. The fermion partners have the same SM quantum numbers as of the right-handed SM fermions, and can therefore mix with the latter after the electroweak and the $U(1)^\prime$ symmetry breaking. As a result, desirable loop-level corrections to the $(g-2)_\mu$, the $W$-boson mass $m_W$ and the Wilson coefficient $\mC_9$ in $b \to s \mu^+ \mu^-$ transitions can be obtained. The final allowed parameter space is also consistent with the constraints from the $Z \to \mu^+ \mu^-$ decay, the neutrino trident production and the LHC direct searches for the vector-like quarks and leptons.}

\begin{document}

\date{}
\maketitle

\section{Introduction}

The Standard Model (SM) of particle physics has proven incredibly successful in describing most phenomena observed in experiments~\cite{Workman:2022ynf}. At present, the major focus of the Large Hadron Colliders (LHC) is the direct searches for new particles and new interactions beyond the SM. While no confirmed direct signals for new physics (NP) beyond the SM have been observed at the LHC so far, several interesting hints of NP have been emerging from the precision measurements.

The long-standing anomaly of the muon $g-2$ provides an intriguing hint of NP. The latest measurement by the Muon $g-2$ collaboration at Fermilab~\cite{Muong-2:2021ojo}, after combined with the previous measurement by the Brookhaven E821 experiment~\cite{Muong-2:2006rrc}, shows a $4.2\,\sigma$ discrepancy with the SM prediction~\cite{Aoyama:2020ynm}:
\begin{align}
\Delta a_{\mu}=a_\mu^{\rm exp}-a_\mu^\SM=(251\pm59)\times10^{-11},
\end{align}
with $a_\mu=(g-2)_\mu/2$. On the theoretical side, further detailed studies are presently on going to improve the precision of the hadronic contribution to the muon $g-2$~\cite{Borsanyi:2020mff,Ce:2022kxy,ExtendedTwistedMass:2022jpw,Blum:2023qou,CMD-3:2023alj} (see also ref.~\cite{Colangelo:2022jxc} for a recent review).

Another interesting hint of NP comes from the updated measurement of the $W$-boson mass by the Collider Detector at Fermilab (CDF) collaboration~\cite{CDF:2022hxs}. Using the complete dataset collected by the CDF II detector, the CDF collaboration reported a value~\cite{CDF:2022hxs}
\begin{align}
  m_W^{\rm CDF}=80.4335 \pm 0.0064_{\rm stat} \pm 0.0069_{\rm syst}\GeV.
\end{align}
Such a high precision measurement deviates from the average of the previous measurements from LEP, CDF, D0 and ATLAS, $m_W^{\rm PDG}=80.379\pm 0.012\GeV$~\cite{Workman:2022ynf}, as well as from the LHCb measurement $m_W^{\rm LHCb} = 80.354 \pm 0.031\GeV$~\cite{LHCb:2021bjt}, and the recent ATLAS measurement $m_W^{\rm ATLAS}=80.360\pm 0.016\GeV$~\cite{ATLAS:2023fsi}. Furthermore, it shows a $7\sigma$ deviation from the SM expectation obtained through a global electroweak (EW) fit, $m_W^{\rm SM}=80.357 \pm 0.006\GeV$~\cite{Workman:2022ynf}. If confirmed by future precision measurements, this anomaly could imply another sign of NP beyond the SM.

Finally, several interesting discrepancies with the SM predictions have been observed in the $b \to s \ell^+ \ell^-$ (with $\ell=e, \mu$) processes over the last decade. Notably, the experimental picture has changed dramatically at the end of 2022: the previous $R_{K^{(*)}}$ anomalies were not confirmed by the updated LHCb measurements~\cite{LHCb:2022qnv,LHCb:2022zom}, and the recent CMS measurement of the branching fraction $\mB(B_s \to \mu^+ \mu^-)$~\cite{CMS:2022mgd} made the current world average~\cite{HeavyFlavorAveragingGroup:2022wzx} in excellent agreement with the SM prediction~\cite{Bobeth:2013uxa,Beneke:2017vpq}. However, several observables in the $b \to s \ell^+ \ell^-$ transitions, especially the angular observable $P_5^\prime$ in $B \to K^{\ast} \mu^+ \mu^-$~\cite{LHCb:2020lmf} as well as the branching ratios of $B \to K^{(\ast)} \mu^+ \mu^-$~\cite{LHCb:2014cxe} and $B_s \to \phi \mu^+ \mu^-$~\cite{LHCb:2021zwz}, still show $2\sim 3\,\sigma$ deviations from the corresponding SM predictions. Furthermore, the recent global fits show that the overall consistency of the current $b \to s \ell^+ \ell^-$ data with the theoretical predictions can be significantly improved by adding NP to the short-distance Wilson coefficient $\mC_9$~\cite{Gubernari:2022hxn,Greljo:2022jac,Ciuchini:2022wbq,Alguero:2023jeh,Wen:2023pfq}.

It is noted that, within the SM, the $(g-2)_\mu$, the $W$-boson mass and the $b \to s \ell^+ \ell^-$ transition all receive significant contributions from the loop diagrams with quarks or leptons. These loop diagrams could also be mediated by the NP fermions that have the same SM quantum numbers as of the SM ones. Therefore, these NP effects could simultaneously explain the data mentioned above. In this paper, in order to investigate such a possibility, we construct a NP model in which the SM is extended with the vector-like fermion partners that are featured by additional $U(1)^\prime$ gauge symmetry. We assume that the fermion partners are $SU(2)_L$ singlets and have the same SM quantum numbers as of the right-handed SM fermions. After the EW and the $U(1)^\prime$ symmetry are spontaneously broken, possible mixings between the fermion partners and the SM right-handed fermions are obtained, which can result in desirable loop-level corrections to the $(g-2)_\mu$, $m_W$, and $\mC_9$. Our model can be regarded as an extension of the one introduced in refs.~\cite{Li:2022gwc,Kamenik:2017tnu}. We will also consider the various constraints coming from the $Z \to \mu^+ \mu^-$ decay, the neutrino trident production and the LHC direct searches for the vector-like quarks and leptons.

The paper is organized as follows: In section~\ref{sec:Zp}, we introduce the NP model based on a new $U(1)^\prime$ symmetry. In section~\ref{sec:theory}, we recapitulate the theoretical framework for the various processes and investigate the NP effects on them. Our detailed numerical results and discussions are presented in section~\ref{sec:numerics}. We conclude in section~\ref{sec:conclusion}. Details of the one-loop corrections to the $Z\mu\mu$ coupling and the global fit of the $b \to s \ell^+ \ell^-$ processes are presented in appendices~\ref{app:Zmumu} and \ref{app:fit}, respectively. The relevant loop functions are collected in appendix~\ref{sec:loop}.

\section{Model}
\label{sec:Zp}

\begin{table}[ht]
	\tabcolsep 0.095in
	\renewcommand{\arraystretch}{1.3}
	\centering
	\begin{tabular}{ @{} l  c c c c c c c c c c c c c c @{} }
		\toprule
		&        \multicolumn{7}{c}{SM}    & \phantom{a}        & \multicolumn{5}{c}{NP}
		\\\cmidrule{2-8} \cmidrule{10-14}
		& $Q_{3L}$ & $u_{3R}$ & $L_{2L}$  & $L_{3L}$ & $e_{2R}$ & $e_{3R}$ & $H$ &    & $U_{L/R}^\prime$ & $E_{L/R}$ & $\Phil$ & $\Phit$ & $\phi$
		\\\midrule
		$SU(3)_C$                 & 3       & 3        & 1        & 1        & 1       & 1        & 1   &   & 3              & 1        & 1           & 1       & 1
		\\
		$SU(2)_L$                 & 2       & 1        & 2        & 2        & 1       & 1        & 2    &  & 1              & 1        & 1           & 1       & 1
		\\
		$\phantom{S}U(1)_Y$       & 1/6     & 2/3      & $-1/2$   & $-1/2$   & $-1$    & $-1$     & $1/2$ & & 2/3            & $-1$     & 0           & 0       & 0
		\\
		$\phantom{S}U(1)^\prime$ & 0       & 0        & $q_\ell$ & $-q_\ell$ & $q_\ell$ & $-q_\ell$& 0     & & $q_t$          & 0        &  $-q_\ell$    & $q_t$   & 0
		\\\bottomrule
	\end{tabular}
	\caption{Quantum numbers of the SM and NP fields in our model under the $SU(3)_C \otimes SU(2)_L \otimes U(1)_Y \otimes U(1)^\prime$ gauge symmetry. See text for more details.} \label{tab:model}
\end{table}

As discussed in ref.~\cite{Li:2022gwc}, in order to simultaneously accommodate the $m_W^{\rm CDF}$, the $(g-2)_{\mu}$ anomaly and the $b \to s \ell^+ \ell^-$ discrepancies, one can introduce the new fermions that are characterized by additional $U(1)^\prime$ gauge symmetry and have the same SM quantum numbers as of the SM ones. To this end, let us firstly introduce our model based on the new $U(1)^\prime$ symmetry, where the SM and NP fields as well as their charges under the $SU(3)_C \otimes SU(2)_L \otimes U(1)_Y \otimes U(1)^\prime$ gauge symmetry are given in table~\ref{tab:model}.

\subsection{Quark sector}

The quark sector of the model is identical to that introduced in refs.~\cite{Li:2022gwc,Kamenik:2017tnu}. All the SM quarks do not carry the  $U(1)^\prime$ charge, while a vector-like top partner with
\begin{align}
U_{L/R}^\prime=\left(\textbf{3} ,\, \textbf{1} ,\, 2/3 ,\, q_t\right),
\end{align}
and a complex scalar with 
\begin{align}
  \Phit=(\textbf{1} ,\, \textbf{1} ,\, 0 ,\, q_t),
\end{align}
are introduced, where the quantum numbers in brackets dictate the transformations under the $SU(3)_C \otimes SU(2)_L \otimes U(1)_Y \otimes U(1)^\prime$ gauge symmetry.

The general Lagrangian involving the top partner $U_{L/R}^\prime$, the SM Higgs doublet $H$ and the scalar $\Phit$ can be written as
\begin{align}\label{eq:Lagrangian:quark}
  \mathcal{L} \supset  q_t g^{\prime} \left(\bar{U}_{L}^{\prime} \gamma^{\mu} U_{L}^{\prime}+\bar{U}_{R}^{\prime} \gamma^{\mu} U_{R}^{\prime}\right) Z_{\mu}^{\prime}
  -\Big(\sum_{i} \lambda_{ii} \bar{Q}_{i L} \tilde{H} u_{i R}+ \lambda_{4i} \bar{U}_{L}^{\prime} u_{i R} \Phit +\mu \bar{U}_{L}^{\prime} U_{R}^{\prime}+\text {h.c.}\Big),
\end{align}
where $g^{\prime}$ denotes the $U(1)^\prime$ gauge coupling, and $Q_{iL}=(u_{iL},d_{iL})^T$ and $u_{iR}$ ($i=1,\,2,\,3$) stand for the $i$-th generation of the left-handed quark doublet and the right-handed quark singlet of the SM, respectively. As discussed in refs.~\cite{Fajfer:2013wca,Aguilar-Saavedra:2013wba}, mixings of $U_{L/R}^\prime$ with the first two generations suffer from severe experimental constraints. As a consequence, such mixings are assumed to be small compared to its mixing with the third generation and can be therefore neglected, as done in refs.~\cite{Kamenik:2017tnu,Fox:2018ldq}; this leads us to set $\lambda_{41}=\lambda_{42}=0$. In the following, we will use the abbreviations $\lambda_H=\lambda_{33}$ and $\lambda_\Phit=\lambda_{43}$.

The fermions present in eq.~(\ref{eq:Lagrangian:quark}) are all given in the interaction eigenbasis. Without loss of generality, we have chosen the basis where the up-type Yukawa matrix is diagonal in the $3\times 3$ SM flavor space. After the EW and the $U(1)^\prime$ symmetry breaking, the mass matrix for the up-type quarks takes the form
\begin{align}\label{eq:mass matrix:quark}
  \begin{pmatrix}
    \lambda_{11}v_H & 0 & 0 & 0 \\
    0 & \lambda_{22} v_H& 0 & 0 \\
    0 & 0& \lambda_H v_H & 0 \\
    0 & 0& \lambda_\Phit v_\Phit& \sqrt 2 \mu
  \end{pmatrix},
\end{align}
where the vacuum expectation values (vevs) of $H$ and $\Phit$ fields are defined by $\langle H \rangle=v_H/\sqrt{2}$ and $\langle \Phit \rangle=v_{\Phit}/\sqrt{2}$, respectively. When diagonalizing the mass matrix, only the rotation between $u_3$ and $U^\prime$ is needed in the up sector, and the CKM matrix is defined by the rotation among the down-type quarks. We refer the readers to refs.~\cite{Fajfer:2013wca,Aguilar-Saavedra:2013wba} for more details. As a result, the physical top quarks $(t_L, t_R)$ and their partners $(T_L, T_R)$ are related to the fermions present in eq.~(\ref{eq:Lagrangian:quark}) through~\cite{Fox:2018ldq}
\begin{align}\label{eq:rotation:q}
\begin{pmatrix}
t_{L} \\
T_{L}
\end{pmatrix}
=R(\theta_L)
\begin{pmatrix}
u_{3 L} \\
U_{L}^{\prime}
\end{pmatrix}
,\qquad\qquad
\begin{pmatrix}
t_{R} \\
T_{R}
\end{pmatrix}
=
R(\theta_R)
\begin{pmatrix}
u_{3 R} \\
U_{R}^{\prime}
\end{pmatrix}
,
\end{align}
with the rotation matrix given by
\begin{align}\label{eq:rotation matrix}
R(\theta)=
\begin{pmatrix}
\cos \theta &-\sin \theta \\
\sin \theta & \hphantom{-}\cos \theta
\end{pmatrix}
,
\end{align}
where the mixing angles $\theta_L$ and $\theta_R$ parameterize the rotation matrices of the left- and right-handed quarks, respectively. In terms of the physical parameters, the top-quark mass $m_t$, the top-partner mass $m_T$ and the two mixing angles are related to each other through
\begin{align}\label{eq:theta_L}
\tan \theta_L=\frac{m_t}{m_T} \tan \theta_R,
\end{align}
with $m_t$ and $m_T$ determined by
\begin{align}
\label{eq:Yukawa:H:q}
\lambda_{H}&=\frac{\cos \theta_{L}}{\cos\theta_R} \frac{\sqrt{2} m_{t} }{v_H}, \quad \quad
  \\
\label{eq:Yukawa:Phit}
\lambda_{\Phit}&= \cos \theta_L\sin \theta_R
\frac{\sqrt{2}   m_T}{v_{\Phit}}\left(1- \frac{m_t^2}{ m_T^2}\right).
\end{align}

In the fermion mass eigenbasis, the gauge interactions involving the top quark and the top partner take the form~\cite{Kamenik:2017tnu,Fox:2018ldq}
\begin{align}
  \label{eq:interaction:q:gamma}
\mathcal L_\gamma^t =&\frac{2}{3} e \bar{t} \slashed{A} t+\frac{2}{3} e \bar{T} \slashed{A} T,
  \\[0.5em]
  \label{eq:interaction:q:W}
\mathcal L_W^t =&\frac{g}{\sqrt{2}}V_{td_i} \left(c_L \bar{t} \slashed{W} P_{L} d_i+s_L \bar{T} \slashed{W} P_{L} d_i\right)+\text { h.c.}\,,
  \\[0.5em]
  \label{eq:interaction:q:Z}
  \mathcal L_Z^t =&\frac{g}{c_W}\left(\bar{t}_{L}, \bar{T}_{L}\right)
\begin{pmatrix}
\frac{1}{2} c_L^2-\frac{2}{3} s_W^2 & \frac{1}{2} s_Lc_L \\
\frac{1}{2} s_Lc_L & \frac{1}{2} s_L^2-\frac{2}{3} s_W^2
\end{pmatrix} \slashed{ Z}
\begin{pmatrix}
t_{L} \\
T_{L}
\end{pmatrix}
   +\frac{g}{c_W}\left(\bar{t}_{R}, \bar{T}_{R}\right) \bigg(-\frac{2}{3}s_W^2 \bigg) \slashed{ Z}
\begin{pmatrix}
t_{R} \\
T_{R}
\end{pmatrix},
  \\[0.5em]
  \label{eq:interaction:q:Zp}
\mathcal L_\Zp^t=&g_t\left(\bar{t}_{L}, \bar{T}_{L}\right)
\begin{pmatrix}
s_L^2 & -s_Lc_L \\
-s_Lc_L & c_L^2
\end{pmatrix}
\slashed{Z}^{\prime}
\begin{pmatrix}
t_{L} \\
T_{L}
\end{pmatrix}
  + (L \to R),
\end{align}
where $g$ is the SM weak coupling constant, $\gt= q_t \gp$, $c_{L,R}=\cos\theta_{L,R}$, $s_{L,R}=\sin\theta_{L,R}$, and $s_W=\sin\theta_W$ with $\theta_W$ being the Weinberg angle; $d_i \in \lbrace d,s,b \rbrace$ denote the down-type quarks and $V_{td_i}$ the CKM matrix elements. In eq. (\ref{eq:interaction:q:W}), $V_{td_i}$ arise from the rotation among the left-handed down-type quarks, while $c_L$ and $s_L$ from the rotation specified by eq.~(\ref{eq:rotation:q}). The scalar interactions involving the top quark and the top partner can be written as
\begin{align}\label{eq:Yukawa}
\mathcal L_h^t &=-\frac{m_{t}}{v_{H}}\left(\bar{t}_{L}, \bar{T}_{L}\right) \left(\begin{array}{cc}
c_L^2 & c_L^2 \tan \theta_{R} \\
s_L c_L & s_L c_L \tan \theta_{R}
\end{array}\right) h\begin{pmatrix}
t_{R} \\
T_{R}
\end{pmatrix}+\text{h.c.}\,,
  \\
  \mathcal L_\Phit&= -\frac{\lambda_{\Phi_t}}{\sqrt{2}}\left(\bar{t}_{L}, \bar{T}_{L}\right) \left(\begin{array}{cc}
-s_L c_R & s_L s_R  \\
c_L c_R & c_L s_R 
\end{array}\right) h\begin{pmatrix}
t_{R} \\
T_{R}
\end{pmatrix}+\text{h.c.}\,.
\end{align}
Except for the interactions involving $\Zp$ and $\Phit$, the quark sector of this model is similar to that of the generic vector-like quark models, which have been extensively studies in the literature~\cite{delAguila:1982fs,Branco:1986my,Langacker:1988ur,Lavoura:1992np,delAguila:2000rc,Okada:2012gy,Dawson:2012di,Aguilar-Saavedra:2013qpa,Ellis:2014dza,Alves:2023ufm}.

\subsection{Lepton sector}

In the lepton sector, the model is based on the $U(1)_{L_{\mu}-L_{\tau}}$ gauge symmetry~\cite{He:1990pn,He:1991qd}. Both the second- and the third-generation leptons are charged under the $U(1)^\prime$ gauge group. Explicitly, their quantum numbers under the $SU(3)_C\otimes SU(2)_L \otimes U(1)_Y\otimes U(1)^\prime$ gauge symmetry are given by
\begin{align}
L_{2L} &=\left(\textbf{1} ,\, \textbf{2} ,\, -1/2 ,\, +q_\ell\right), & e_{2R}&=\left(\textbf{1} ,\, \textbf{1} ,\, -1 ,\, +q_\ell\right),
\\
L_{3L}&=\left(\textbf{1} ,\, \textbf{2} ,\, -1/2 ,\, -q_\ell\right), & e_{3R}&=\left(\textbf{1} ,\, \textbf{1} ,\, -1 ,\, -q_\ell\right),
\nonumber
\end{align}
where $L_{2L}$ and $e_{2R}$ ($L_{3L}$ and $e_{3R}$) denote the second-generation (third-generation) left-handed lepton doublet and right-handed lepton singlet of the SM, respectively. $q_\ell$ denotes the charge of the $U(1)^\prime$ gauge symmetry. We also introduce a vector-like muon partner
\begin{align}
E_{L/R}=\left(\textbf{1} ,\, \textbf{1} ,\, -1 ,\, 0\right),
\end{align}
as well as two complex scalar fields
\begin{align}
\phi=(\textbf{1} ,\, \textbf{1} ,\, 0 ,\, 0), \quad\quad  \Phi_\ell=\left(\textbf{1} ,\, \textbf{1} ,\, 0 ,\, -q_\ell\right).
\end{align}
After spontaneous symmetry breaking, the field $\phi$ provides mass to the muon partner, while the field $\Phi_\ell$ induces mixing between the muon lepton and the muon partner.

The general Lagrangian involving the $\Zp$ boson and the scalars $H$, $\Phil$ and $\phi$ is given by
\begin{align}\label{eq:Lagrangian:lepton}
\mathcal L \supset &\,q_\ell g^{\prime} \left(\bar{L}_{2L} \gamma^{\mu} L_{2L}+\bar{e}_{2R} \gamma^{\mu} e_{2R}-\bar{L}_{3L} \gamma^{\mu} L_{3L}-\bar{e}_{3R} \gamma^{\mu} e_{3R}\right) Z_{\mu}^{\prime}
  \\
  &\, -\Big[ \sum _{i} \lambda_{ii}^\ell \bar{L}_{iL} H e_{i R}+ \lambda_{42}^\ell \bar E_L e_{2 R} \Phil + \lambda_{43}^\ell \bar E_L e_{3R} \Phi_\ell^*
    +\big(\lambda_{41}^\ell \bar E_L e_{1R}+\lambda_{44}^\ell \bar E_L E_R \big) \phi+\text {h.c.} \Big],
    \nonumber
\end{align}
where $L_{iL}=(\nu_{iL}, e_{iL})^T$ and $e_{iR}$ ($i=1,\, 2,\, 3$) are the $i$-th generation of the SM left-handed lepton doublet and right-handed lepton singlet, respectively. Here, for simplicity, we assume that the mixings of $E_{L/R}$ with the first and third generations are small and can be therefore neglected, \textit{i.e.}, $\lambda_{43}^\ell=\lambda_{41}^\ell=0$. The abbreviations $\eta_H=\lambda_{22}^\ell$, $\lambda_\Phil=\lambda_{42}^\ell$ and $\lambda_\phi=\lambda_{44}^\ell$ will be used in the following.

In eq.~(\ref{eq:Lagrangian:lepton}), the fermions are all given in the interaction eigenbasis. After the EW and the $U(1)^\prime$ symmetry breaking, where the scalars $H$, $\Phil$ and $\phi$ acquire their vevs, the resulting mass matrix for the charged leptons is similar to eq.~(\ref{eq:mass matrix:quark}). After diagonalization, the physical muons $(\mu_L, \mu_R)$ and their partners $(M_L, M_R)$ can be expressed as
\begin{align}\label{eq:rotation:l}
\begin{pmatrix}
\mu_{L} \\
M_{L}
\end{pmatrix}
=
R(\delta_L)
\begin{pmatrix}
e_{2 L} \\
E_{L}
\end{pmatrix}
,
\qquad\qquad
\begin{pmatrix}
\mu_{R} \\
M_{R}
\end{pmatrix}
=
R(\delta_R)
\begin{pmatrix}
e_{2 R} \\
E_{R}
\end{pmatrix}
,
\end{align}
where the mixing angles $\delta_L$ and $\delta_R$ parameterize the rotation matrices of the left- and right-handed leptons, respectively. The two mixing angles are related to each other through
\begin{align}\label{eq:eta_L}
\tan \delta_{L}=\frac{m_{\mu}}{m_{M}} \tan \delta_R,
\end{align}
where $m_\mu$ and $m_M$ denote the muon and the muon-partner mass, respectively. Together with the mixing angles and the vevs of $H$, $\Phi_\ell$ and $\phi$, the physical masses $m_\mu$ and $m_M$ can be expressed in terms of the Yukawa couplings $\eta_{H}$, $\lambda_{\Phi_\ell}$ and $\lambda_{\phi}$ as
\begin{align}
m_{\mu}&=\frac{1}{\sqrt{2}}\big(\cos\delta_L \cos \delta_{R} v_H \eta_{H}-\sin\delta_L \cos \delta_{R} v_{\Phi_\ell}\lambda_{\Phi_\ell}+\sin\delta_L \sin \delta_{R}v_{\phi}\lambda_{\phi}\big),
  \\
m_{M}&=\frac{1}{\sqrt{2}}\big(\sin\delta_L \sin \delta_{R} v_H \eta_{H}+\cos\delta_L \sin \delta_{R} v_{\Phi_\ell}\lambda_{\Phi_\ell}+\cos\delta_L \cos \delta_{R}v_{\phi}\lambda_{\phi}\big),
\end{align}
and vice versa as
\begin{align}
\eta_{H}&=\frac{\cos \delta_{L}}{\cos\delta_R} \frac{\sqrt{2} m_{\mu} }{v_H}, \label{eq:Yukawa:H:l}
\\
\lambda_{\Phi_\ell}&=   \cos \delta_{L}\sin \delta_{R} 
\frac{\sqrt{2}   m_{M}}{v_{\Phi_\ell}}\left(1- \frac{m_{\mu}^2}{ m_{M}^{2}}\right), \label{eq:Yukawa:Phil}
\\
\lambda_{\phi}&=\frac{\cos \delta_{R}}{\cos\delta_L} \frac{\sqrt{2} m_{M} }{v_{\phi}}, \label{eq:Yukawa:phi} 
\end{align}
where $v_{\Phi_\ell}=\sqrt 2 \langle \Phi_\ell \rangle$, $v_\phi=\sqrt 2 \langle \phi \rangle$, and $v_H$ is the vev of the SM Higgs doublet.

In the fermion mass eigenbasis, explicit expressions of the gauge interactions involving the muon and the muon partner can be written as
\begin{align}
  \mathcal L_\gamma^\ell =&- e \bar{\mu} \slashed{A} \mu- e \bar{M} \slashed{A} M,
                            \label{eq:interaction:l:gamma}
  \\[0.5em]
\mathcal L_W^\ell =&\frac{g}{\sqrt{2}} \left(\hat{c}_L \bar{\mu} \slashed{W} P_{L} \nu_{\mu}+\hat{s}_L \bar{M} \slashed{W} P_{L} \nu_{\mu}\right)+\text { h.c.}\,,
                     \label{eq:interaction:l:W}
  \\[0.5em]
\mathcal L_Z^\ell =&\frac{g}{c_W}\left(\bar{\mu}_{L}, \bar{M}_{L}\right)
\begin{pmatrix}
-\frac{1}{2} \hat{c}_L^2+ s_W^2 & -\frac{1}{2} \hat{s}_L\hat{c}_L \\
-\frac{1}{2} \hat{s}_L\hat{c}_L & -\frac{1}{2} \hat{s}_L^2+s_W^2
\end{pmatrix} \slashed{ Z}
\begin{pmatrix}
\mu_{L} \\
M_{L}
\end{pmatrix}
 +\frac{g}{c_W} s_W^2\left(\bar{\mu}_{R}, \bar{M}_{R}\right)  \slashed{ Z}
\begin{pmatrix}
\mu_{R} \\
M_{R}
\end{pmatrix},
  \label{eq:interaction:l:Z}
  \\[0.5em]
\mathcal L_\Zp^\ell=&\left[g_\ell \left(\bar{\mu}_{L}, \bar{M}_{L}\right)
\begin{pmatrix}
\hat{c}_L^2 & \hat{s}_L \hat{c}_L \\
\hat{s}_L \hat{c}_L & \hat{s}_L^2
\end{pmatrix}
\slashed{Z}^{\prime}
\begin{pmatrix}
\mu_{L} \\
M_{L}
\end{pmatrix}
  -g_\ell  \bar\tau_L \slashed\Zp \tau_L
  + (L \to R)\right]
  \label{eq:interaction:l:Zp}
  \\
  & \qquad + g_\ell  \left(\bar{\nu}_{\mu} \slashed{Z}^{\prime} P_{L} \nu_{\mu}-\bar{\nu}_{\tau} \slashed{Z}^{\prime} P_{L} \nu_{\tau}\right), \nonumber
\end{align}
where $\gl = q_\ell \gp$, $\hat{s}_{L,R}=\sin \delta_{L,R}$ and $\hat{c}_{L,R}=\cos\delta_{L,R}$. The Yukawa interactions of the scalars with the muon and the muon partner are given by
\begin{align}\label{eq:interaction:l:phi}
\mathcal L_h&=-\frac{m_{\mu}}{v_{H}}\left(\bar{\mu}_{L}, \bar{M}_{L}\right) \left(\begin{array}{cc}
\hat{c}_L^2 & \hat{c}_L^2 \tan \delta_{R} \\
\hat{s}_L \hat{c}_L & \hat{s}_L \hat{c}_L \tan \delta_{R}
\end{array}\right) h\begin{pmatrix}
\mu_{R} \\
M_{R}
\end{pmatrix}+\text{h.c.}\,,\\
\mathcal L_{\Phi_\ell}&=-\frac{\lambda_{\Phi_\ell}}{\sqrt{2}}\left(\bar{\mu}_{L}, \bar{M}_{L}\right) \left(\begin{array}{cc}
-\hat{s}_L \hat{c}_R & -\hat{s}_L \hat{s}_R \\
\hat{c}_L \hat{c}_R & \hat{c}_L \hat{s}_R
\end{array}\right)\Phi_\ell\begin{pmatrix}
\mu_{R} \\
M_{R}
\end{pmatrix}+\text{h.c.}\,, \\
\mathcal
L_{\phi}&=-\frac{\lambda_{\phi}}{\sqrt{2}}\left(\bar{\mu}_{L}, \bar{M}_{L}\right) \left(\begin{array}{cc}
\hat{s}_L \hat{s}_R & -\hat{s}_L \hat{c}_R \\
-\hat{c}_L \hat{s}_R & \hat{c}_L \hat{c}_R
\end{array}\right)\phi\begin{pmatrix}
\mu_{R} \\
M_{R}
\end{pmatrix}+\text{h.c.}\,,
\end{align}
where $h$, $\Phil$ and $\phi$ denote the physical scalar fields after spontaneous symmetry breaking.

\subsection{Choice of the model parameters}

To complete our model, we also need to specify the scalar potential, which is given by
\begin{align}
  \mathcal V = & \sum_S \left[ \mu_S^2 |S|^2 + \text{Re}\big(\lambda_S^{(3)} \phi \big) |S|^2 - \lambda_S^{(4)} |S|^4\right]
  \\
      &+\big( \lambda_{H\phi}|\phi|^2 + \lambda_{H\Phit} |\Phit|^2  + \lambda_{H\Phil}|\Phi_{\ell}|^2 \big) H^{\dag} H
        + \big( \lambda_{\phi\Phit} |\Phit|^2 + \lambda_{\phi\Phil} |\Phi_{\ell}|^2 \big)|\phi|^2  + \lambda_{\Phit\Phil} |\Phit|^2|\Phi_{\ell}|^2,
        \nonumber
\end{align}
where $S\in\{H, \phi, \Phit, \Phil\}$. The potential is assumed to be such that all the scalar fields acquire only real vevs. The scalar fields $\Phit$ and $\Phil$ are responsible for the spontaneous breaking of the $U(1)^\prime$ gauge symmetry, and give mass to the $\Zp$ boson, with $m_\Zp^2={g^{\prime}}^2 (v_{\Phit}^2+v_\Phil^2)$.

In this paper, we focus on the parameter region with $m_T>m_t$ and $m_M > m_\mu$. In the quark sector, without loss of generality, the parameters $\lambda_H$, $v_H$ and $\mu$ are chosen to be positive. After taking into account the relations in eqs.~(\ref{eq:Yukawa:H:q}) and (\ref{eq:Yukawa:Phit}), the cases with $v_\Phit >0$ and $v_\Phit<0$ correspond to the ranges of $0\leqslant\theta_L<\theta_R<\pi/2$ and $\pi/2<\theta_R<\theta_L\leqslant\pi$, respectively. Similarly, in the lepton sector, the parameters $\eta_H$, $\lambda_{\Phi_{\ell}}$ and $\lambda_{\phi}$ are also chosen to be positive. Considering the relations in eqs.~\eqref{eq:Yukawa:H:l}--\eqref{eq:Yukawa:phi}, we find that $v_\phi>0$, and the cases with $v_\Phil >0$ and $v_\Phil<0$ correspond to the ranges of $0\leqslant\delta_L<\delta_R<\pi/2$ and $\pi/2<\delta_R<\delta_L\leqslant\pi$, respectively. In the following analysis, we will focus on the case where all the vevs are positive and the mixing angles are restricted within the regions of $0\leqslant\theta_L<\theta_R<\pi/2$ and $0\leqslant\delta_L<\delta_R<\pi/2$.

\section{Theoretical framework}
\label{sec:theory}

In this section, we investigate the relevant observables affected by our model, including the $W$-boson mass, the $(g-2)_\mu$, the $b \to s \ell^+ \ell^-$ processes, as well as the $Z\mu^+\mu^-$ couplings. The NP contributions to most of them arise at the loop level. During our evaluations, all the loop diagrams are calculated in the unitary gauge. As in ref.~\cite{Li:2022gwc}, the computations are implemented in two independent methods by using different packages including \texttt{FeynRules}~\cite{Alloul:2013bka}, \texttt{FeynArts}~\cite{Hahn:2000kx}, \texttt{FeynCalc}~\cite{Mertig:1990an,Shtabovenko:2016sxi,Shtabovenko:2020gxv}, \texttt{Package-X}~\cite{Patel:2016fam}, as well as some in-house routines.

\subsection{$W$-mass shift and oblique parameters}
\label{sec: EWPT}

\begin{figure}[t]
  \centering
  \subfloat[\label{fig:vp:a}]{\includegraphics[width=0.25\textwidth]{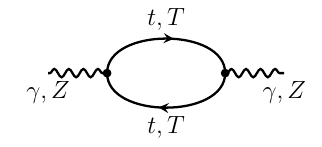}}
  \subfloat[\label{fig:vp:b}]{\includegraphics[width=0.25\textwidth]{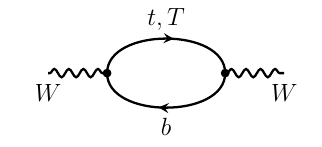}}
  \subfloat[\label{fig:vp:c}]{\includegraphics[width=0.25\textwidth]{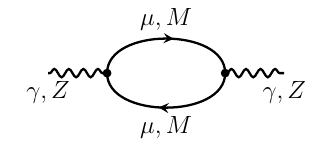}}
  \subfloat[\label{fig:vp:d}]{\includegraphics[width=0.25\textwidth]{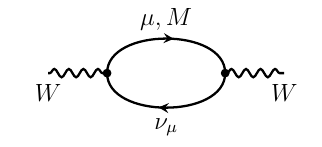}}
  \caption{One-loop Feynman diagrams for the vacuum polarizations of gauge bosons in our model.}
  \label{fig:vp}
\end{figure}

The global fit to the EW precision data, known as the global EW fit~\cite{Hollik:1988ii,Langacker:1991zr,Erler:2019hds}, is a powerful tool to test the SM as well as to probe possible NP effects~\cite{Flacher:2008zq,Baak:2014ora,Haller:2018nnx,deBlas:2021wap}. As an important EW precision observable, the $W$-boson mass in the SM is determined from the global EW fit. Recently, the CDF $m_W$ measurement~\cite{CDF:2022hxs} shows large deviation from the SM prediction, which could be explained by NP contributions; see \textit{e.g.} refs.~\cite{Balkin:2022glu,Endo:2022kiw,Babu:2022pdn,Strumia:2022qkt,Asadi:2022xiy,Gu:2022htv,Lu:2022bgw,deBlas:2022hdk} and the references therein. In our model, the $W$-boson mass shift can be divided into the following three parts:
\begin{align}\label{eq:deltamw}
 \Delta m_W = \Delta m_W^Q+\Delta m_W^L + \Delta m_W^{W\mu \nu},
\end{align}
where $\Delta m_W^Q$, $\Delta m_W^L$ and $\Delta m_W^{W\mu \nu}$ denote the contributions from the top-partner, the muon-partner and the modified $W\mu\nu$ vertex, respectively. We show in figure~\ref{fig:vp} the relevant one-loop Feynman diagrams for these NP contributions.

The oblique parameters $S$, $T$ and $U$ encode most of the NP effects on the SM EW sector~\cite{Peskin:1991sw,Peskin:1990zt,Maksymyk:1993zm}. To be specific, these parameters capture the NP contributions from the gauge-boson vacuum-polarization corrections and can be generically written as~\cite{Peskin:1991sw,Peskin:1990zt}
\begin{align}
S&=\frac{4 s_W^2c_W^2}{\alpha_e } \bigg[\frac{\Pi_{Z Z}\left(m_Z^2\right)- \Pi_{Z Z}(0)}{m_Z^2}-\frac{c_W^2-s_W^2}{s_W c_W}  \Pi_{Z \gamma}^{\prime}(0)- \Pi_{\gamma \gamma}^{\prime}(0)\bigg],
\nn\\
T &=\frac{1}{\alpha_e}\bigg[\frac{ \Pi_{W W}(0)}{ m_W^2}-\frac{ \Pi_{Z Z}(0)}{m_Z^2}\bigg],
\nn\\
U &=\frac{4 s_W^2}{\alpha_e }\bigg[\frac{\Pi_{W W}\left(m_W^2\right)- \Pi_{W W}(0)}{m_W^2}-\frac{c_W}{s_W}\Pi_{Z\gamma}^\prime(0)- \Pi_{\gamma \gamma}^\prime(0)\bigg]-S, \nn
\end{align}
where $\Pi_{XY}$ denotes the vacuum polarization of the gauge fields with $X,Y=W,Z,\gamma$, $c_W=\cos\theta_W$, and $\alpha_e$ is the fine structure constant. In terms of these oblique parameters, the $W$-boson mass shift can be written as~\cite{Peskin:1991sw}
\begin{align}
\Delta m_W^2=\frac{\alpha_e c_W^2 m_Z^2}{c_W^2-s_W^2}\left[-\frac{S}{2}+c_W^2 T+\frac{c_W^2-s_W^2}{4 s_W^2} U\right].
\end{align}
Therefore, to explain the discrepancy of the CDF $m_W$ measurement from the SM expectation, we need a global EW fit to the oblique parameters $S$, $T$ and $U$, which could be affected by contributions from both the quark and the lepton sector in our model. Let us discuss them in turn.

In the model introduced in section~\ref{sec:Zp}, extra contributions to the vacuum polarizations of gauge fields arise from the diagrams shown in figure~\ref{fig:vp}. Figures~\ref{fig:vp:a} and \ref{fig:vp:b} encode the contributions from the modified quark-gauge couplings that are characterized by the mixing angle $\theta_L$ and the loops involving the top partner, respectively. Their contributions to the oblique parameters read~\cite{Li:2022gwc}
\begin{align}
S_Q &= \frac{s_L^2}{12\pi} \Big[ K_1(y_t, y_T) + 3 c_L^2 K_2 (y_t, y_T) \Big], \label{eq:S} \\[0.2cm]
T_Q &=  \frac{3s_L^2}{16\pi s_W^2} \bigg[x_T-x_t - c_L^2\Big(x_T+x_t +\frac{2x_tx_T}{x_T-x_t}\ln\frac{x_t}{x_T} \Big)\bigg], \label{eq:T} \\[0.2cm]
U_Q &= \frac{s_L^2}{12\pi} \Big[ K_3(x_t,y_t)-K_3(x_T,y_T)\Big] -S_Q, \label{eq:U}
\end{align}
where $x_q=m_q^2/m_W^2$ and $y_q=m_q^2/m_Z^2$ for $q=t,T$. Explicit expressions of the loop functions $K_{1,2,3}(x,y)$ are recapitulated in appendix~\ref{sec:loop}.

Similar to the quark sector, contributions from the lepton sector arise from figures~\ref{fig:vp:c} and \ref{fig:vp:d}. Their contributions to the oblique parameters can be written as
\begin{align}
S_L &=\frac{\hat{s}_L^2}{12\pi} \Big[ K_4(y_{\mu}, y_M) +  \hat{c}_L^2 K_5 (y_{\mu}, y_M) \Big], \label{ss} \\[0.2cm]
T_L &=\frac{\hat{s}_L^2}{16\pi s_W^2} \bigg[x_M-x_{\mu} - \hat{c}_L^2\Big(x_M+x_{\mu} +\frac{2x_{\mu}x_M}{x_M-x_{\mu}}\ln\frac{x_{\mu}}{x_M} \Big) \Big], \label{tt}\\[0.2cm]
U_L &= \frac{\hat{s}_L^2}{12\pi} \Big[ K_6(x_{\mu},y_{\mu})-K_6(x_M,y_M)\Big] -S_L,\label{uu}
\end{align}
where $x_\ell=m_\ell^2/m_W^2$ and $y_\ell=m_\ell^2/m_Z^2$ for $\ell=\mu,M$, $\hat{s}_L=\sin \delta_L$ and $\hat{c}_L=\cos \delta_L$. Explicit expressions of the loop functions $K_{4,5,6}(x,y)$ are listed in appendix~\ref{sec:loop}. From eqs.~\eqref{ss}--\eqref{uu} (eqs.~\eqref{eq:S}--\eqref{eq:U}), one can see that the contributions to the oblique parameters $S$, $T$ and $U$ from the lepton (quark) sector are solely determined by the two NP parameters $\delta_L$ and $m_M$ ($\theta_L$ and $m_T$) and are proportional to $\sin^2\delta_L$ ($\sin^2\theta_L$).

Finally, let us discuss the last term of eq.~\eqref{eq:deltamw}, $\Delta m_W^{W\mu \nu}$, which is induced by the modified $W\mu\nu$ coupling and characterized by the mixing angle $\delta_L$, as given in eq.~\eqref{eq:interaction:l:W}. The modified $W\mu\nu$ coupling can affect the muon lifetime, from which the Fermi constant $G_F$ is extracted. During the global EW fit, the prediction for $m_W$ is obtained from its relation to $G_F$. This implies that the input $G_F$ used for calculating $m_W$ should be the modified one rather than the one given by the Particle Data Group (PDG)~\cite{Workman:2022ynf}. Therefore, the NP correction to the $W\mu\nu$ coupling indirectly translates to a shift of the $W$-boson mass, $\Delta m_W^{W\mu \nu}$, which can be written as~\cite{Domingo:2011uf,Heinemeyer:2006px,Awramik:2003rn}
\begin{align}\label{eq:MW}
\frac{(m_W^{\text{SM}}+\Delta m_W^{W\mu \nu})^2}{m_Z^2}=\frac{1}{2}+\sqrt{\frac{1}{4}-\frac{\pi \alpha_e}{\sqrt{2} G_{F} m_Z^2}|1+\Delta r|},
\end{align}
where $\Delta r$ encodes the NP correction to the muon lifetime. Specific to our model, the modified $W\mu \nu$ coupling results in $\Delta r=\cos \delta_L-1$.

\subsection{$b \to s \ell^+ \ell^-$ transitions}
\label{bsll}

\begin{figure}[t]
  \centering
  \includegraphics[width=0.241\linewidth]{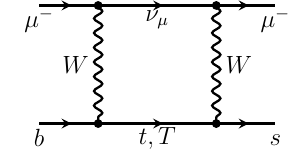}
  \includegraphics[width=0.241\linewidth]{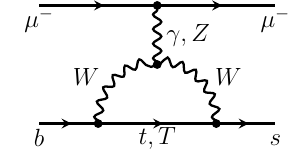}
  \includegraphics[width=0.241\linewidth]{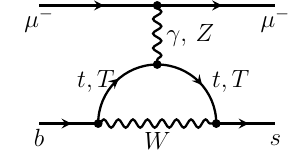}
  \includegraphics[width=0.241\linewidth]{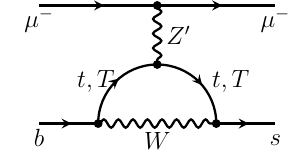}
  \caption{One-loop Feynman diagrams for the $b \to s \mu^+\mu^-$ transitions, including the $\Zp$ and $T$ contributions, as well as the effects of the modified $Z\mu\mu$, $W\mu\nu$ and $Wtb$ couplings.}
  \label{fig:b2s:diagram}
\end{figure}

Our model can also be efficiently explored through the quark-level $b \to s \ell^+ \ell^-$ transitions, such as the $B \to K^{(\ast)} \ell^+ \ell^- $, $B_s \to \ell^+ \ell^-$ and $B_s \to \phi \ell^+ \ell^-$ decays. Here the NP contributions arise firstly at the one-loop level, with the relevant Feynman diagrams shown in figure~\ref{fig:b2s:diagram}. As found already in ref.~\cite{Li:2022gwc}, the NP contributes only to the short-distance Wilson coefficients $\mC_9$ and $\mC_{10}$ of the low-energy effective Hamiltonian governing the $b \to s \ell^+ \ell^-$ transitions~\cite{Buchalla:1995vs}
\begin{align}
\mathcal{H}_{\text{eff}}\supset-\frac{4 G_F}{\sqrt{2}} V_{t b} V_{t s}^{*} \frac{\alpha_e}{4 \pi}\left(\mC_{9\ell} \mathcal{O}_{9\ell}+ \mC_{10\ell} \mathcal{O}_{10\ell}\right)+\text {h.c.},
\end{align}
with the two semi-leptonic operators given by $\mathcal{O}_{9\ell}=\left(\bar{s} \gamma^{\mu} P_L b\right)\left(\bar\ell \gamma_{\mu} \ell\right)$ and $\mathcal{O}_{10\ell}=\left(\bar{s} \gamma^{\mu} P_L b\right)\left(\bar\ell \gamma_{\mu} \gamma_5 \ell\right)$, respectively. The NP contributions to $\mC_{9\ell}$ and $\mC_{10\ell}$ can be divided into the Lepton-Flavor Universal (LFU) and the Lepton-Flavor Violating (LFV) parts. The former arise from the diagrams with the SM gauge bosons, and are given by
\begin{align}\label{eq:WC}
  \mC_{9\ell}^{\rm NP}=&s_L^2 I_1+ s_L^2 \left(1-\frac{1}{4s_W^2}\right) \left(I_2 + c_L^2 I_3 \right),
  \\[0.15cm]
  \mC_{10\ell}^{\rm NP}=& \frac{s_L^2}{4s_W^2}\left(I_2 + c_L^2 I_3 \right),
\end{align}
with $s_{L,R}=\sin \theta_{L,R}$ and $c_{L,R} = \cos \theta_{L,R}$. They are proportional to $\sin^2\theta_L$. The LFV contributions can be written as
\begin{align}
 \mC_{9\mu}^{\rm NP}=\Delta\mC_+^{W,Z}+\Delta\mC_+^\Zp,\\
 \mC_{10\mu}^{\rm NP}=\Delta\mC_-^{W,Z}+\Delta\mC_-^\Zp,
\end{align}
with
\begin{align}\label{eq:WC:Zp}
  \Delta\mC_{\pm}^{W,Z}=&\pm\frac{\hat{s}_L^2 s_L^2}{8s_W^2} (I_6+c_L^2 I_7),\\
 \Delta\mC_{\pm}^\Zp =& \frac{(\hat{c}_R^2\pm \hat{c}_L^2)g_\ell g_t }{e^2} \frac{m_W^2}{m_\Zp^2}c_L^2s_R^2 \left( I_4 -\frac{c_L^2}{c_R^2} I_5 \right),
\end{align}
where $\gl = q_\ell \gp$, $\gt = q_t \gp$, $\hat{s}_{L,R} = \sin \delta_{L,R}$ and $\hat{c}_{L,R} = \cos \delta_{L,R}$. Here the contributions $\Delta\mC_{\pm}^{W,Z}$ arise from the diagrams involving the $W$ and $Z$ bosons, while $\Delta\mC_{\pm}^\Zp$ from the $\Zp$-penguin diagrams. In our case of small mixing angles $\theta_L$ and $\delta_L$, as will be demonstrated in section~\ref{sec:numerics}, $\Delta\mC_{\pm}^{W,Z}$ will be highly suppressed because they are proportional to $\hat{s}_L^2 s_L^2=\sin^2\delta_L\sin^2\theta_L$. Thus, we can safely take the approximations $\mC_{9e}^{\rm NP}\approx\mC_{10e}^{\rm NP}\approx0$ and $\big(\mC_{9\mu}^\NP,\, \mC_{10\mu}^\NP\big) \approx \big(\Delta\mC_+^\Zp,\, \Delta\mC_-^\Zp\big)$. The loop integrals $I_{1-7}$ are functions of $m_{t,T}$ and $m_{\mu,M}$, whose explicit expressions are given in appendix~\ref{sec:loop}. Keeping only the leading terms in $\sin\theta_R$ and $m_W^2/m_T^2$, our results are in agreement with that obtained in ref.~\cite{Fox:2018ldq}.

\subsection{Muon $g-2$}
\label{sec: g-2}

\begin{figure}[t]
  \centering
  \includegraphics[width=0.241\linewidth]{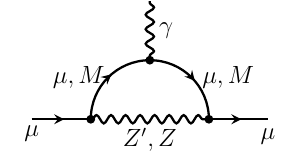}\quad
  \includegraphics[width=0.241\linewidth]{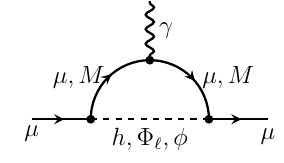}\quad
  \includegraphics[width=0.241\linewidth]{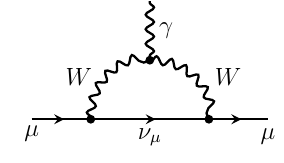}
  \caption{One-loop Feynman diagrams for the muon $(g-2)$, including the $\Zp$, $\Phil$ and $\phi$ contributions, as well as the effects of the modified $Z\mu\mu$, $h\mu\mu$ and $W\mu\nu$ couplings.}
  \label{fig:g-2:diagram}
\end{figure}

The muon anomalous magnetic moment $a_{\mu}\equiv
(g-2)_{\mu}/2$ can provide very promising probes of potential NP effects~\cite{Aoyama:2020ynm,Jegerlehner:2009ry}. Specific to our model, the observable $a_\mu$ is affected by the one-loop Feynman diagrams shown in figure~\ref{fig:g-2:diagram}, which involve the $\Zp$, $\phi$ and $\Phil$ bosons, as well as the SM diagrams but with the modified $Z\mu\mu$, $h\mu\mu$ and $W\mu\nu$ couplings. Their total contributions to $\Delta a_{\mu}$ can be written as
\begin{align}\label{eq:deltaamu}
\Delta a_\mu=\Delta a_\mu^\Zp+\Delta a_\mu^\phi+\Delta a_\mu^\Phil+\Delta a_\mu^W +\Delta a_\mu^Z +\Delta a_\mu^h,
\end{align}
where $\Delta a_\mu^i=\Delta a_\mu^{i,\, \mu}+\Delta a_\mu^{i,\, M}$ ($i=Z,\, \Zp,\, \phi,\, \Phil$), and $\Delta a^{i,\, j}_\mu$ ($j=\mu,\, M$) denotes the contribution from the diagrams involving the particles $i$ and $j$. Their explicit expressions are given, respectively, by
\begin{align}
  \Delta a_\mu^{Z,\,\mu}=&-\frac{g^2 w_{\mu}(\hat{c}_L^4-2\hat{s}_L^2s_W^2-1)}{48\pi^2c_W^2},
                           \label{g-2-1}\\
  \Delta a_\mu^{Z,\,M}=&-\frac{\hat{c}_L^2\hat{s}_L^2w_{\mu}g^2(5w_M^4-14w_M^3+39w_M^2-38w_M+8-18w_M^2 \ln w_M)}{384\pi^2c_W^2(w_M-1)^4},
                         \label{g-2-0}\\
  \Delta a_\mu^{\Zp,\,\mu}=&-\frac{  g_\ell^2 x_{\mu}\left(\hat{c}_L^4-3 \hat{c}_L^2 \hat{c}_R^2+\hat{c}_R^4\right)}{12  \pi^2},
                             \label{g-21}\\
  \Delta a_\mu^{\Zp,\,M}=&\frac{\hat{s}_L^2 \hat{c}_R^2 g_\ell^2 x_M \left(x _M^3+3 x_M-4-6 x_M \ln x_M\right)}{16 \pi^2\left(x_M-1\right)^{3}},
                           \label{g-22}\\
  \Delta a_\mu^{\phi,\,\mu}=&\frac{ \hat{s}_L^2 \hat{s}_R^2 \lambda_{\phi}^2 }{32\pi^2} f_s(z_\mu),
                              \label{g-24}\\
  \Delta a_\mu^{\phi,\,M}=&\frac{\hat{s}_L^2 \hat{c}_R^2 \lambda_{\phi}^2 z_M(z_M^2-4z_M+3+2 \ln z_M)}{32\pi^2(z_M-1)^3}
                            \nonumber\\
                         &+\frac{z_{\mu}(\hat{c}_R^2 \hat{s}_L^2+\hat{c}_L^2 \hat{s}_R^2)\lambda_{\phi}^2( z_{\mu}^3-6z_{\mu}^2+3z_{\mu}+2+6x_M \ln z_{\mu})}{192\pi^2(z_M-1)^4},
                           \label{g-23}\\
  \Delta a_\mu^{\Phi_{\ell},\,\mu}=&\frac{ \hat{s}_L^2 \hat{c}_R^2 \lambda_{\Phi_{\ell}}^2 }{32\pi^2}f_s(y_\mu),\label{g-26}\\
  \Delta a_\mu^{\Phi_{\ell},\,M}=&-\frac{\hat{s}_L^2 \hat{c}_R^2 \lambda_{\Phi_{\ell}}^2 y_M (y_M^2-4y_M+3+2 \ln{y_M}) }{32\pi^2(y_M-1)^3 },\label{g-25}\\
  \Delta a_\mu^{W}=&-\frac{5\hat{s}_L^2g^2w_{\mu}}{96\pi^2c_W^2},\label{g-27}
\end{align}
with 
\begin{align}
  f_s(x)=\frac{ 3x -2}{x}  +\frac{(3x-1)}{x^2} \ln x +\frac{2 \sqrt{1-4 x}(x-1)}{x^2} \ln \frac{1+\sqrt{1-4x}}{2\sqrt x },  
\end{align}
where $w_{i}=m_{i}^2/m_Z^2$, $x_{i}=m_{i}^2/m_\Zp^2$, $y_{i}=m_{i}^2/m_{\Phi_{\ell}}^2$ and $z_{i}=m_{i}^2/m_{\phi}^2$ ($i=\mu,\, M$). From these results, one can see that the NP contributions to $a_\mu$ will vanish in the limit of $\delta_L\to 0$ or $\delta_R \to 0$.

\subsection{Neutrino trident production}
\label{sec:NuTP}

\begin{figure}[t]
  \centering
  \includegraphics[width=0.3\linewidth]{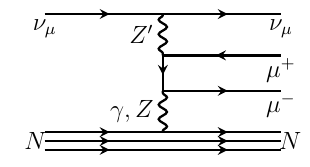}
  \caption{The $\Zp$ contribution to the neutrino trident production in our model. The diagram with $\mu^+$ and $\mu^-$ reversed is not shown.}
  \label{fig:nuTP}
\end{figure}

The $\Zp$ couplings to muons are significantly constrained by the rare process of neutrino trident production, \textit{i.e.}, the production of a $\mu^+\mu^-$ pair from the scattering of a muon-neutrino with heavy nucleus~\cite{CHARM-II:1990dvf,CCFR:1991lpl,NuTeV:1998khj,NuTeV:1999wlw}. In our model, the $\Zp$ boson contributes to this rare process through the tree-level Feynman diagram shown in figure~\ref{fig:nuTP}. The ratio of the cross section in our model to that in the SM, $r_{\nu \rm TP} \equiv \sigma_\NP/\sigma_\SM$, is calculated to be~\cite{Fox:2018ldq,Altmannshofer:2014pba,Altmannshofer:2014cfa}
 \begin{align}\label{ntpf}
r_{\nu \rm TP}=\frac{\left|g_\ell^2 \left( \hat{c}_L^2+ \hat{c}_R^2\right)/m_\Zp^2+\sqrt{2}\left(1+4 s_W^2\right) G_F\right|^2+\left|g_\ell^2\left(\hat{c}_L^2- \hat{c}_R^2\right)/m_\Zp^2+\sqrt{2} G_F\right|^2}{2 G_F^2\left[1+\left(1+4 s_W^2\right)^2\right]},
 \end{align}
where $\gl=g_\ell \gp$ and $\hat c_{L,R}=\cos\delta_{L,R}$.

\subsection{$Z\mu \mu$ and $W\mu \nu$ couplings}
\label{sec:zclep}

\begin{figure}[t]
  \centering
  \includegraphics[width=0.191\linewidth]{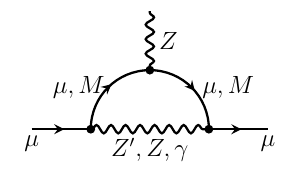}
  \includegraphics[width=0.191\linewidth]{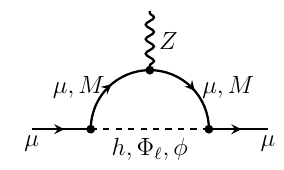}
  \includegraphics[width=0.191\linewidth]{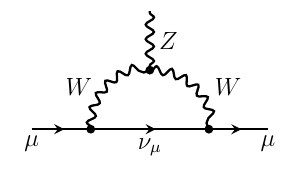}
  \includegraphics[width=0.191\linewidth]{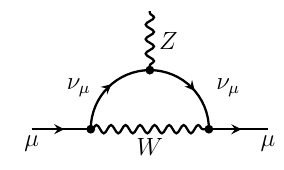}
  \includegraphics[width=0.191\linewidth]{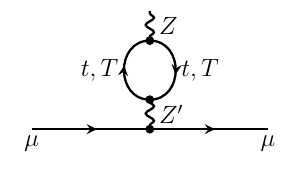}
  \caption{Relevant Feynman diagrams for the one-loop corrections to the $Z\mu\mu$ vertex in our model.}
  \label{fig:diagram:Zmumu}
\end{figure}

Any modification of the $Z$ couplings to leptons must receive stringent constraints from the LEP measurements at the $Z$ pole~\cite{ALEPH:2005ab}. In our model, the effective $Z$ couplings to leptons are given by
\begin{align}\label{eq:zmumulr}
\mathcal L=\frac{g}{ c_W} \bar{\ell}\slashed{Z}(g_{L\ell}P_L+g_{R\ell}P_R)\ell\,,
\end{align}
with $\ell=\mu$ or $e$. The effective couplings $g_{L\mu/R\mu}$ include both the tree-level couplings specified by eq.~(\ref{eq:interaction:l:Z}) and the one-loop vertex corrections shown in figure~\ref{fig:diagram:Zmumu}. The latter can be written as
\begin{align}
  \Delta g_{\Gamma\mu} = \Delta g_{\Gamma\mu}^\Zp + \Delta g_{\Gamma\mu}^\Phil + \Delta g_{\Gamma\mu}^\phi + \Delta g_{\Gamma\mu}^W + \Delta g_{\Gamma\mu}^Z + \Delta g_{\Gamma\mu}^\gamma + \Delta g_{\Gamma\mu}^h  + \Delta g_{\Gamma\mu}^{Z-\Zp}\,,
\end{align}
with $\Gamma=L$ or $R$. Here $\Delta g_{\Gamma\mu}^i$ denotes the correction from the diagram involving the particle $i$ for $i=\Zp$, $\Phil$, $\phi$, $W$, $Z$, $\gamma$ and $h$. As done in ref.~\cite{Denner:1991kt}, these corrections are calculated in the on-shell renormalization scheme. Furthermore, we find that the renormalization of the mixing angle $\delta_L$ has to be taken into account. Details of the calculation are given in appendix~\ref{app:Zmumu}. The vertex corrections $\Delta g_{\Gamma\mu}^{\Zp,\,\phi,\,\Phil}$ depend on $\gl$, $\lambda_{\Phi_\ell}$, $\lambda_\phi$, $m_{\mu/M}$, $m_\Zp$, $m_{\Phi_\ell}$, $m_\phi$, and their explicit expressions are provided as an ancillary notebook file. In addition, the $Z-\Zp$ mixing via the top (top-partner) loops could also affect the $Z\mu\mu$ couplings~\cite{Li:2022gwc,Camargo-Molina:2018cwu,Dobrescu:2021vak}. Such an effect, denoted by $\Delta g_{\Gamma \mu}^{Z-\Zp}$ in eq.~(\ref{eq:zmumulr}), can be written as
\begin{align}\label{eq:g-2:mixing}
\Delta g_{\Gamma \mu}^{Z-\Zp}=&\frac{3 \gt \gl s_L^2 \hat{c}_\Gamma ^2}{32 \pi^2 (m_Z^2-m_\Zp^2)} (c_L^2I_L + c_R^2I_R)\,,
\end{align}
where the loop functions $I_{L,R}$ depend on $m_t$ and $m_T$, and their explicit expressions can be found in appendix~\ref{sec:loop}. Finally, the contributions involving the SM particles are all proportional to $\sin^2\delta_L$. In our numerical analysis, we will consider the case of small mixing angle $\delta_L$ to avoid large modification of the $Z\mu\mu$ vertex at the tree level. Therefore, these contributions can be safely neglected, which means that $\Delta g_{\Gamma \mu}^{W,Z,\gamma,h}\approx 0$.

To constrain the $Z\mu \mu$ couplings, we follow refs.~\cite{Camargo-Molina:2018cwu,Arnan:2019olv} and consider the following two observables. The first one is the LFU ratio, $R^Z_{\mu/e} \equiv \Gamma(Z\to\mu^+\mu^-)/\Gamma(Z\to e^+ e^-)$, measured at LEP-I~\cite{ALEPH:2005ab}. Using the effective couplings defined in eq.~(\ref{eq:zmumulr}), we can write the decay width of $Z \to \ell^+ \ell^-$ as
\begin{align}
\Gamma(Z\to \ell^+ \ell^-) = \dfrac{m_Z \lambda^{1/2}_Z}{6\pi v^2} \bigg{[}&\left(|g_{L\ell}|^2+|g_{R\ell}|^2\right)\bigg{(}1- \dfrac{m_{\ell}^2}{m_Z^2}\bigg{)}+ 6 \dfrac{m_{\ell}^2}{m_Z^2}\, \mathrm{Re}\big(g_{L\ell}  g_{R\ell}^*\big) \bigg{]}\,,
\end{align}
with $\lambda_Z\equiv m_Z^2(m_Z^2-4m_{\ell}^2)$. The second observable is the leptonic asymmetry parameter defined by~\cite{Camargo-Molina:2018cwu}
\begin{align}
  \mathcal{A}_{\mu}=\frac{\Gamma(Z\to\mu^+_L\mu^-_L)-\Gamma(Z\to\mu^+_R\mu^-_R)}{\Gamma(Z\to \mu^+ \mu^-)},
\end{align}
where $\mu_{L/R}$ corresponds to the left/right-handed muon. Specific to our model, we obtain
\begin{align}
 \Gamma(Z\to\ell^+_\Gamma \ell^-_\Gamma )=&\dfrac{m_Z \lambda^{1/2}_Z}{6\pi v^2}|g_{\Gamma \ell}|^2\bigg (1- \dfrac{m_{\ell}^2}{m_Z^2}\bigg ) \,,
\end{align}
for $\Gamma =L$ or $R$.

From eq.~(\ref{eq:interaction:l:W}), one can see that the mixing angle $\delta_L$ can change the $W\mu\nu$ coupling, and thus affects the branching ratio of the $W^+\to \mu^+ \nu$ decay. Explicitly, we have the relation
\begin{align}\label{eq:Wmunu_Br}
 \frac{\mB(W^+\to\mu^+\nu)}{\mB_\SM(W^+\to\mu^+\nu)}=\cos^2 \delta_L,
\end{align}
which is valid at the tree level.

\section{Numerical analysis}
\label{sec:numerics}

In this section, we proceed to present our numerical results and discussions. Firstly we list in table~\ref{tab:input} the main input parameters used in our numerical analysis.

\begin{table}[t]
	\tabcolsep 0.15in
	\renewcommand{\arraystretch}{1.2}
	\centering
	\begin{tabular}{c c c c}
		\toprule
		Input & Value & Unit & Ref.
		\\\midrule
		$m_t^{\text{pole}}$ & $172.69 \pm 0.30$& $\text{GeV}$ & \cite{Workman:2022ynf}
		\\\midrule
		$s_W^2$ & $0.231 21$& &  \cite{Workman:2022ynf}
		\\
		$G_F$ & $1.166 378 8 \times 10^{-5} $ &$\text{GeV}^{-2}$ &  \cite{Workman:2022ynf}
		\\
		$\alpha_e(m_Z)$ & $1/127.940(14)$
		& & \cite{Workman:2022ynf}
		\\
		$\alpha_{s}(m_Z)$ & 0.11843(81) & & \cite{FlavourLatticeAveragingGroupFLAG:2021npn}
		\\\midrule
		$|V_{cb}|$ & $(41.15 \pm 0.34 \pm 0.45) \times 10^{-3}$& &  \cite{CKMfitter}
		\\
		$|V_{ub}|$ & $(3.88 \pm 0.08 \pm 0.21) \times 10^{-3} $ & &  \cite{CKMfitter}
		\\
		$|V_{us}|f^{K \to \pi}_+(0)$ & $0.2165 \pm 0.0004 $ & & \cite{Workman:2022ynf}
		\\
		$\gamma$&  $72.1^{+5.4}_{-5.7} $ & ${}^\circ$ & \cite{CKMfitter}
		\\
		$f^{K \to \pi}_+(0)$  & $0.9675 \pm 0.0009 \pm 0.0023$ &  & \cite{CKMfitter}
		\\\bottomrule
	\end{tabular}
	\caption{Main input parameters used in our numerical analysis.}
	\label{tab:input}
\end{table}

As discussed in section~\ref{sec:theory}, the relevant independent parameters in our model contain, besides the masses of the NP particles, the mixing angles and the couplings $g_t, g_\ell, \lambda_\phi, \lambda_\Phit, \lambda_\Phil$, where the re-definitions $\gl = q_\ell \gp$ and $\gt = q_t \gp$ have been taken. In our numerical analysis, we consider the following parameter space:
\begin{align}\label{eq:range}
  0 < g_t, g_\ell,\, \lambda_\phi, \lambda_\Phit, \lambda_\Phil < 2.
\end{align}
Taking into account the relations specified by eqs.~\eqref{eq:theta_L} and \eqref{eq:eta_L}, we will take $\theta_L$ and $\delta_L$ as the two independent mixing angles in the quark and lepton sectors, respectively. Their values are varied within the following ranges:
\begin{align}
  0<\sin\theta_L<0.5, \qquad\qquad 0<\sin\delta_L<0.01,
\end{align}
to avoid large tree-level modifications to the $W$, $Z$ and $h$ couplings to fermions, as will be discussed in detail in the next subsection.

\subsection{$Z\mu\mu$ and $W\mu\nu$ couplings}
\label{sec:Zmumu}

As discussed in section~\ref{sec:zclep}, the $Z\mu\mu$ coupling is stringently constrained by the LEP measurements~\cite{ALEPH:2005ab}, especially by the observables $R^Z_{\mu/e}$ and $\mathcal{A}_{\mu}$. In our model, the $Z\mu \mu$ coupling is affected by the mixing angle $\delta_L$ at the tree level, and also receives contributions from the $Z-\Zp$ mixing and the vertex corrections at the one-loop level.

\begin{figure}[t]
  \centering
  \includegraphics[width=0.32\textwidth]{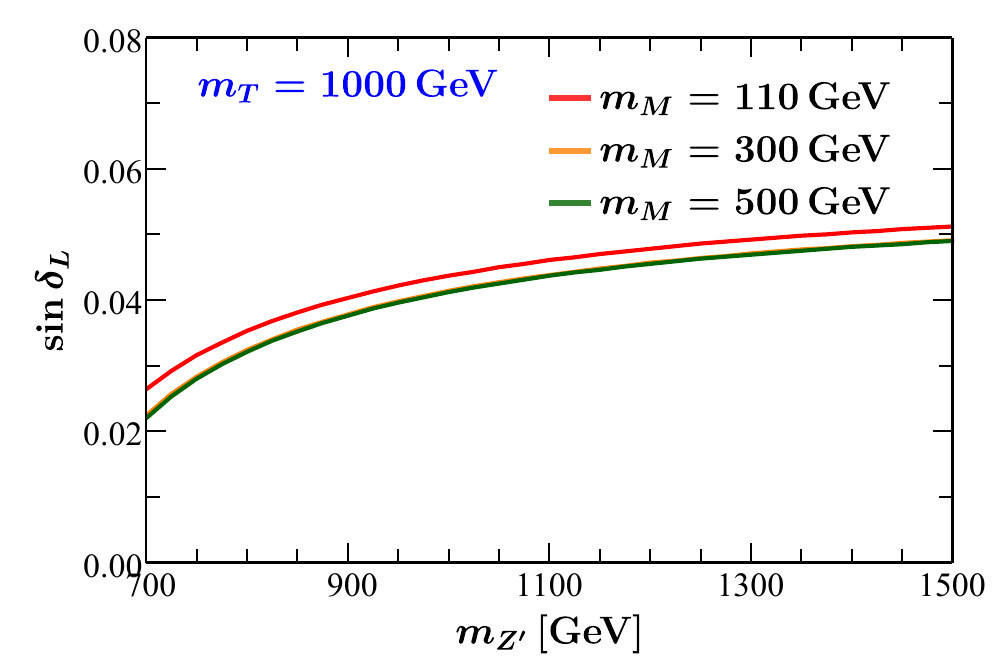}
  \includegraphics[width=0.32\textwidth]{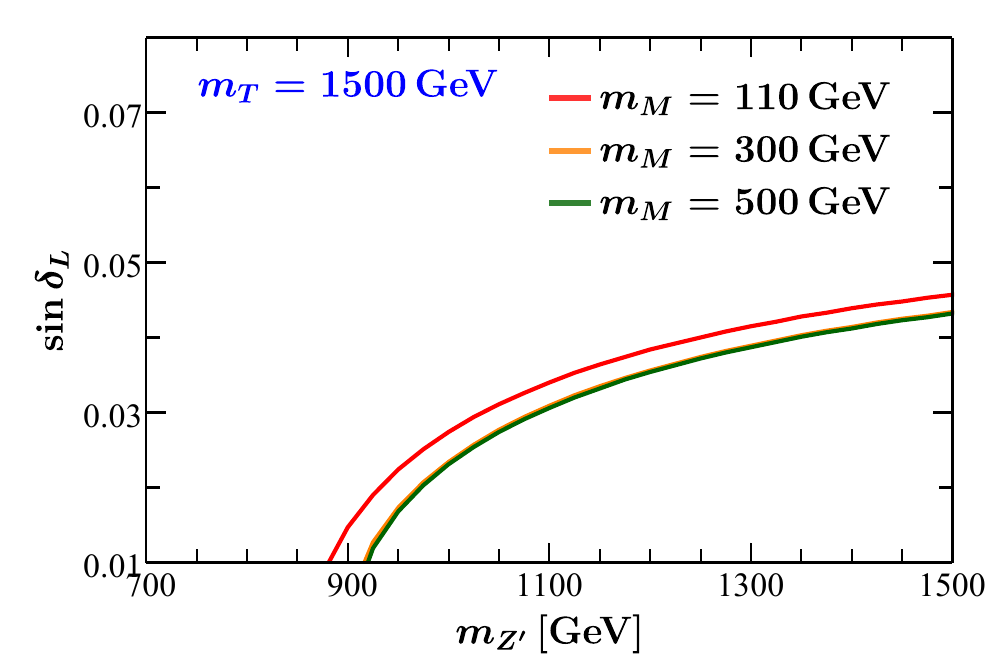}
  \includegraphics[width=0.32\textwidth]{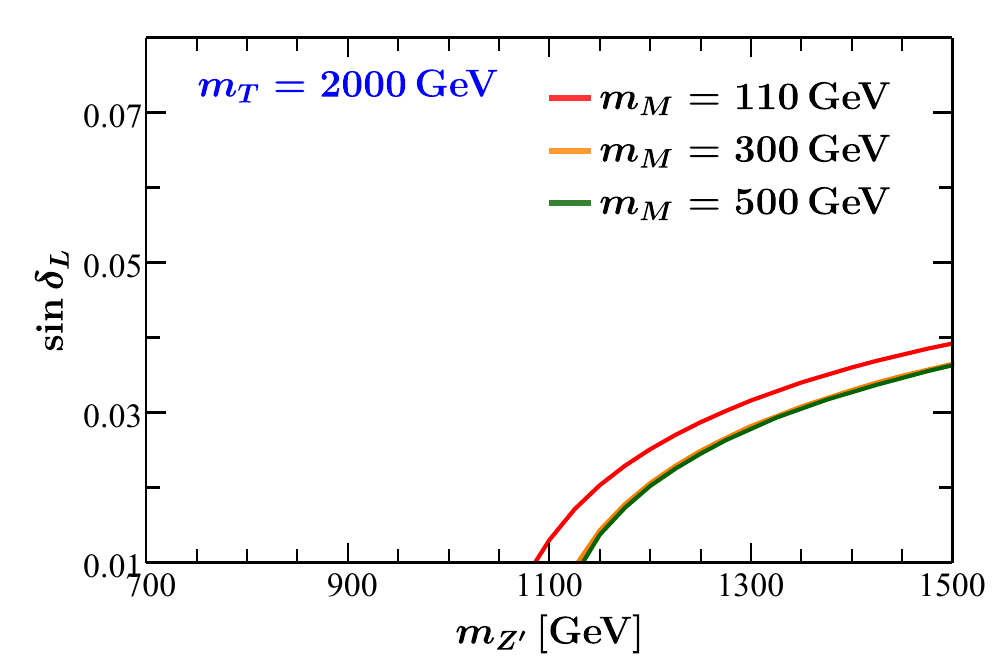}
  \caption{The $2\,\sigma$ upper bounds on $\sin \delta_L$ as a function of $m_\Zp$ from the measured $Z\mu\mu$ coupling, for $m_M=110$, $300$ and $500\GeV$ as well as $m_T=1000$, $1500$ and $2000\GeV$. }
  \label{fig:zmumu}
\end{figure}

In our numerical analysis, we take $\gl=\gt=1$, $\lambda_\phi=\lambda_\Phil=1$,\footnote{We have checked that varying $\lambda_\phi$ and $\lambda_\Phil$ within the range of eq.~(\ref{eq:range}) only changes our results slightly.} $m_\phi=1\GeV$ and $m_\Phil=2\TeV$, which are all consistent with the constraints discussed in the following subsections. The parameter space of $\sin\theta_L$ is chosen within the range allowed by the CDF $m_W$ measurement, as will be derived in section~\ref{mwmass}. Then, the remaining relevant parameters are $(m_\Zp , m_T, m_M, \sin \delta_L)$. Taking the LEP measurements $R^Z_{\mu/e}= 1.0001 \pm 0.0024$~\cite{Workman:2022ynf} and $\mathcal{A}_{\mu}=0.1456\pm 0.0091$~\cite{ALEPH:2005ab} as constraints, we can finally obtain the allowed values of these parameters. As an illustration, we show in figure~\ref{fig:zmumu} the $2\,\sigma$ upper bounds on $\sin \delta_L$ as a function of $m_\Zp$ from the measured $Z\mu\mu$ coupling, for $m_M=110$, $300$ and $500\GeV$ as well as $m_T=1000$, $1500$ and $2000\GeV$. It can be seen that a lighter $\Zp$ or a heavier $T$ implies a stronger upper bound on $\sin \delta_L$, while the upper bound is not sensitive to $m_M$. Numerically, we obtain $\sin \delta_L < 0.05$ in the $2\,\sigma$ allowed parameter space. Furthermore, we have checked that, in the $2\,\sigma$ allowed parameter space, the NP contributions to the $Z\mu\mu$ coupling from the $Z-\Zp$ mixing and the one-loop vertex corrections are both less than $1\%$ of the SM prediction. This in turn means that the fine-tuning among the different NP contributions is small.

On the other hand, the mixing angle $\delta_L$ receives also constraint from the measured branching ratio of $W^+\to\mu^+\nu$ decay, as indicated by eq.~\eqref{eq:Wmunu_Br}. Taking as input the LEP measurement $\mB(W^+\to\mu^+\nu)_{\rm exp}=0.1063(15)$~\cite{Workman:2022ynf} and the SM prediction $\mathcal{B}(W^+\to\mu^+\nu)_{\rm SM}=0.1083$~\cite{Altarelli:1996gh}, we obtain the upper bound $\sin\delta_L<0.43$, which is much weaker than that from the $Z\mu\mu$ coupling.

In the following, we will consider the upper bound $\sin\delta_L < 0.01$, which definitely satisfies the constraint from the $Z\mu\mu$ coupling. In addition, since the $Z\mu\mu$ coupling is not sensitive to $m_M$, a value of $m_M=300\GeV$ will be taken for simplicity. As will be demonstrated later, such a choice of $\sin\delta_L$ and $m_M$ is enough to explain the muon $g-2$ anomaly and the $b\to s \ell^+\ell^-$ discrepancies, while satisfying most of the relevant constraints mentioned in section~\ref{sec:theory}.

\subsection{$W$-boson mass and global EW fit}
\label{mwmass}

\begin{figure}[t]
	\centering
	\includegraphics[width=0.45\textwidth]{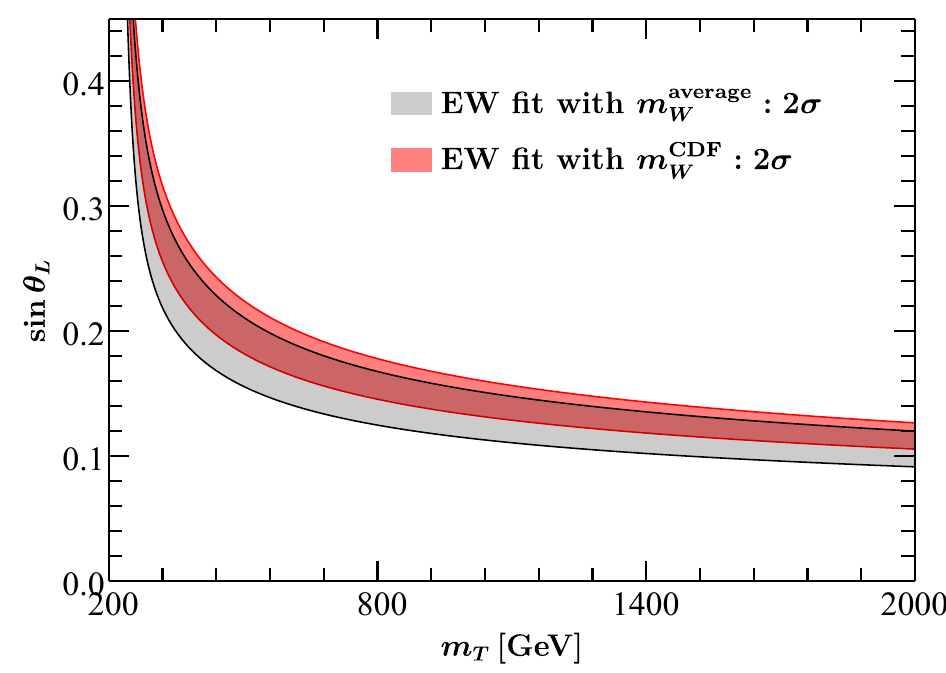}
	\caption{The $2\,\sigma$ allowed regions in the $(m_T, \sin\theta_L)$ plane from the global EW fit. The red and the black region are obtained from the latest CDF measurement $m_W^{\rm CDF}$~\cite{CDF:2022hxs} and from the average of $m_W$~\cite{deBlas:2022hdk} including both $m_W^{\rm CDF}$ and all the previous measurements~\cite{Workman:2022ynf}, respectively.}
	\label{fig:EWPT}
\end{figure}

As the NP effects could affect all the three oblique parameters $S$, $T$ and $U$, a global EW fit is necessary to see if our model could explain the CDF $m_W$ measurement~\cite{CDF:2022hxs}. Recently, the global EW fit including the latest CDF $m_W$ measurement has been performed by several groups~\cite{Strumia:2022qkt,Asadi:2022xiy,Gu:2022htv,Lu:2022bgw,deBlas:2022hdk}. Here we will adopt the result obtained in ref.~\cite{Lu:2022bgw}, which is based on the package \texttt{Gfitter}~\cite{Flacher:2008zq,Baak:2014ora,Haller:2018nnx}.\footnote{Itis found that using the global EW fit results~\cite{deBlas:2022hdk} based on the package \texttt{HEPfit}~\cite{DeBlas:2019ehy} does not substantially change our numerical results.} The resulting values of the oblique parameters, together with their correlations, read~\cite{Lu:2022bgw}
\begin{align}
\label{corre}
\begin{array}{l}
S=0.06 \pm 0.10,\\
T=0.11 \pm 0.12,\\
U=0.14 \pm 0.09,
\end{array}
\qquad\qquad
{\rm cor=}
\begin{bmatrix}
1.00 & 0.90 & -0.59\\
& 1.00 & -0.85 \\
& & \hphantom{-}1.00
\end{bmatrix},
\end{align}
where ``cor" denotes the correlation matrix.

As discussed in section~\ref{sec: EWPT}, the NP contributions arise mainly from the diagrams involving the top-partner, the muon-partner and the modified $W\mu\nu$ coupling. In the following, we will consider the region of $\sin \delta_L < 0.01$ obtained in the last subsection, and investigate these three contributions one by one.

\begin{itemize}
\item By using the input parameters in table~\ref{tab:input} and considering $\sin \delta_L < 0.01$, we find that the $W$-boson mass shift from the modified $W\mu \nu$ coupling (cf. eq.~\eqref{eq:MW}) is given numerically by $\Delta m_W^{W\mu\nu} < 0.007\GeV$. Therefore, such an effect is too small to explain the latest CDF $m_W$ measurement~\cite{CDF:2022hxs}, and can be safely neglected.

\item From eqs.~\eqref{ss}--\eqref{uu}, we can see that the muon-partner contributions to the oblique parameters $S$, $T$ and $U$ are proportional to $\sin^2 \delta_L$. After considering $\sin \delta_L < 0.01$, the resulting shift $\Delta m_W^M$ is highly suppressed and found to be $\mO(10^{-3})$ smaller than from the top-partner contribution. Therefore, we could also safely neglect the muon-partner contributions.

\item The top-partner effects on the $S$, $T$ and $U$ parameters depend only on the mixing angle $\theta_L$ and the top-partner mass $m_T$ (cf. eqs.~(\ref{eq:S})--(\ref{eq:U})). Here we consider $\sin\theta_L<0.5$ to avoid large modification to the $Wtb$ coupling. After taking into account the global EW fit results in eq.~(\ref{corre}), we find that there still exist allowed parameter regions at the $2\,\sigma$ level, as shown in figure~\ref{fig:EWPT}. It can be inferred that $\sin\theta_L=0.14\sim0.20$ within the mass range $500<m_T<2000\GeV$. 
\end{itemize}

Consequently, the latest CDF $W$-boson mass shift can be explained in our model, and the allowed parameter regions shown in figure~\ref{fig:EWPT} will be used in the following numerical analysis. When the parameters vary within the regions, deviations of the top-Higgs coupling (cf. eq.~\eqref{eq:Yukawa}) from its SM value are less than $10\%$, which are also consistent with the current Higgs measurements at the LHC~\cite{Fox:2011qd,Fox:2018ldq,CMS:2018uag,ATLAS:2020qdt}. In addition, for comparison, we show in figure~\ref{fig:EWPT} the allowed parameter regions by taking as constraint the average of $m_W$~\cite{deBlas:2022hdk} including both $m_W^{\rm CDF}$ and all the previous measurements~\cite{Workman:2022ynf}. In this case, the allowed values of $\sin\theta_L$ become smaller.

\subsection{Muon $g-2$ and neutrino trident production}
\label{g-2g-2ntp}

The $U(1)_{L_{\mu}-L_{\tau}}$ extension of the SM can explain the $(g-2)_\mu$ anomaly, but receives significant constraint from the rare process of neutrino trident production~\cite{Altmannshofer:2014pba}. In this subsection, we will investigate the possibility of explaining the $(g-2)_\mu$ anomaly in our model, while satisfying the constraint from neutrino trident production. This rare process has been searched for in several neutrino beam experiments, including CHARM-II~\cite{CHARM-II:1990dvf}, CCFR~\cite{CCFR:1991lpl} and NuTeV~\cite{NuTeV:1998khj,NuTeV:1999wlw}. Combining the data from these collaborators, the ratio of the measured cross section to that in the SM is given by $\sigma_{\exp}/\sigma_{\mathrm{SM}}=0.95 \pm 0.25$~\cite{Altmannshofer:2014cfa}, which should be confronted with the theoretical result in eq.~(\ref{ntpf}).

\begin{figure}[t]
	\centering
	\includegraphics[width=0.45\textwidth]{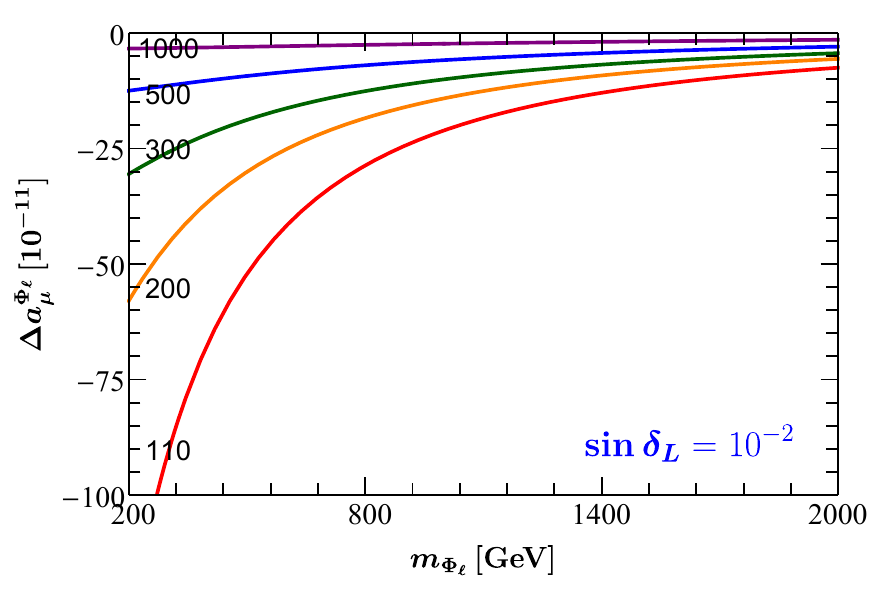}
	\qquad
	\includegraphics[width=0.45\textwidth]{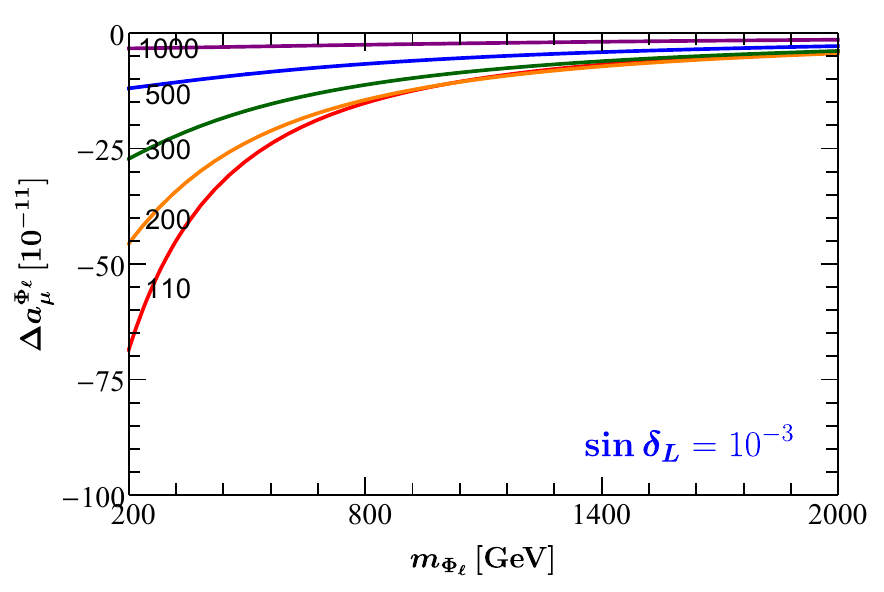}
	\caption{The $\Phi_{\ell}$ contribution to $\Delta a_{\mu}$ as a function of $m_{\Phi_{\ell}}$ for $\sin \delta_L=10^{-2}$ (left) and $10^{-3}$ (right), with $m_M=110$, $200$, $300$, $500$ and $1000\GeV$.}
	\label{fig:phil}
\end{figure}

\begin{figure}[t]
	\centering
	\includegraphics[width=0.32\textwidth]{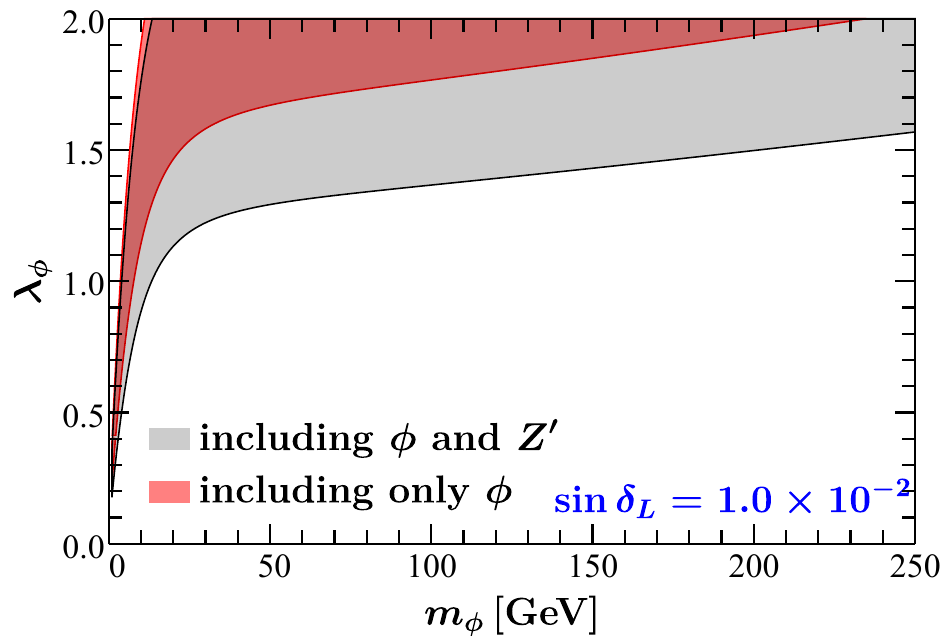}
	\includegraphics[width=0.32\textwidth]{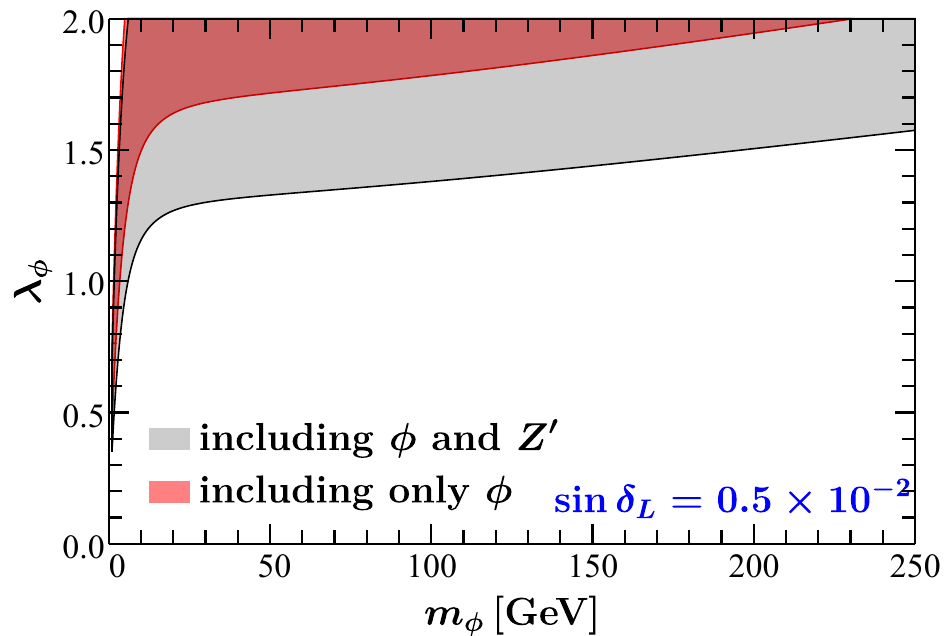}
	\includegraphics[width=0.32\textwidth]{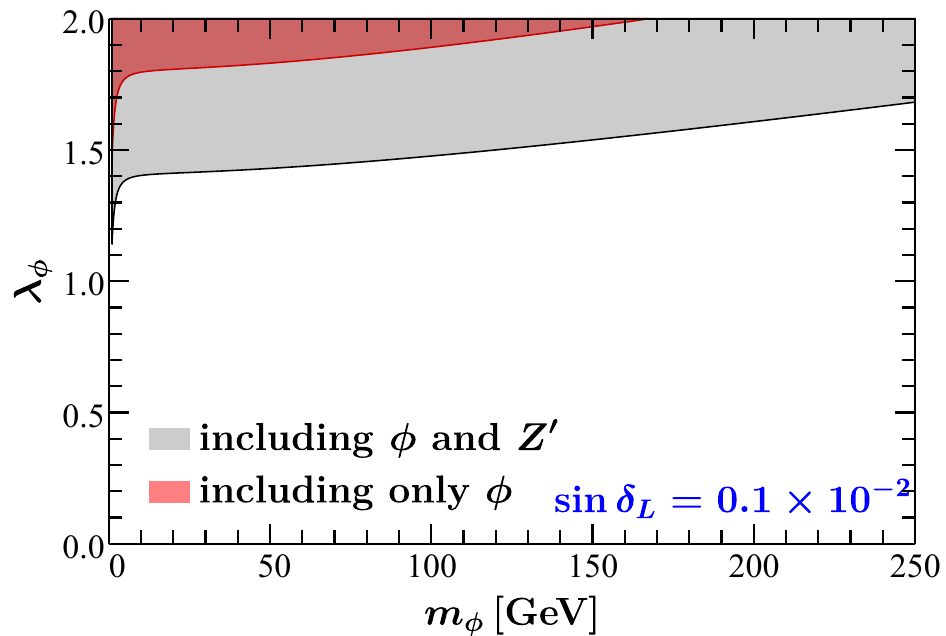}
	\caption{Constraints on $(m_{\phi}, \lambda_{\phi})$ from the $(g-2)_{\mu}$ for $\sin \delta_L=(1.0,\, 0.5,\,0.1) \times 10^{-2}$. The red and the gray region denote the $2\,\sigma$ allowed parameter space including only the $\phi$ and both the $\phi$ and $\Zp$ contributions, respectively. For the $\Zp$ contribution, $m_\Zp=1\TeV$ and $\gl=2$ are taken, and the constraint from neutrino trident production has been taken into account.}
	\label{fig:g-2lamphi-mphi}
\end{figure}

\begin{figure}[t]
	\centering
	\includegraphics[width=0.45\textwidth]{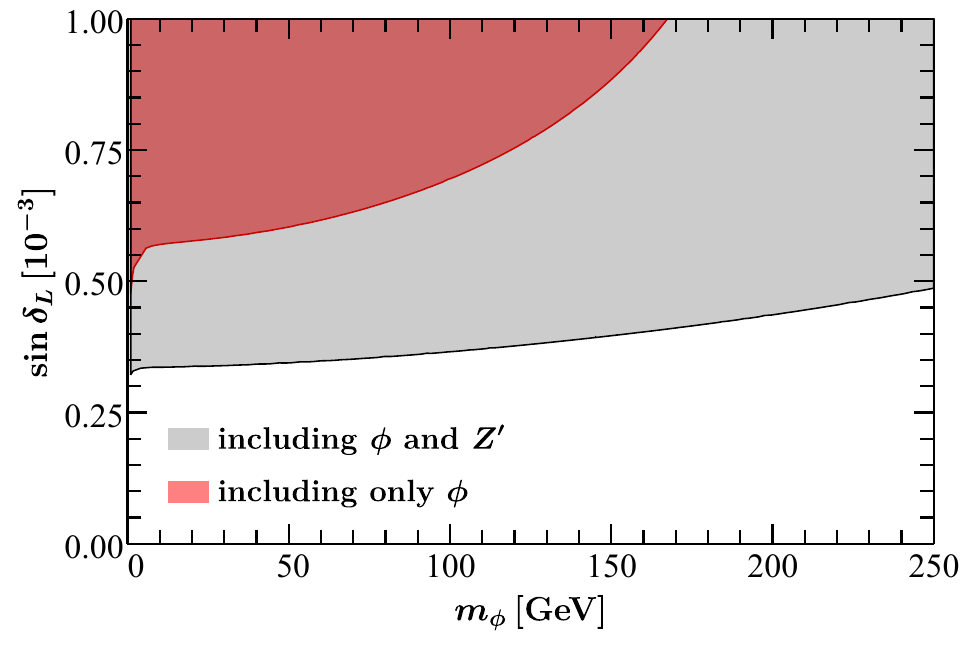}
	\qquad
	\includegraphics[width=0.45\textwidth]{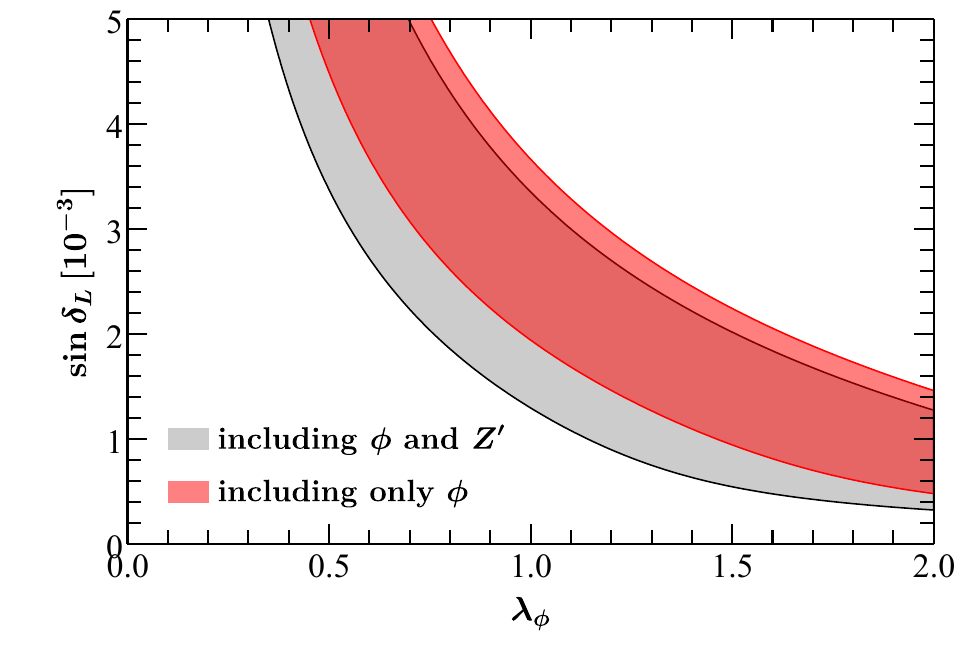}
	\caption{Constraints on $(\sin \delta_L, m_{\phi}, \lambda_{\phi})$ from the $(g-2)_{\mu}$ at the $2\,\sigma$ level, which are plotted in the $(m_{\phi},\, \sin \delta_L)$ plane for $\lambda_\phi=2$ (left) and in the $(\lambda_\phi,\, \sin\delta_L)$ plane for $m_\phi=1\GeV$. The color captions are the same as in figure~\ref{fig:g-2lamphi-mphi}.}
	\label{fig:g-2ssl-mphi}
\end{figure}

The NP contributions to $(g-2)_\mu$ can be divided into three parts (cf. eq.~(\ref{eq:deltaamu})), arising from the penguin diagrams involving $\Zp$, $\Phi_{\ell}$ and $\phi$, respectively. Let us now discuss them one by one.

\begin{itemize}

\item From eqs.~\eqref{g-26}--\eqref{g-25}, one can see that the $\Phil$ contribution $\Delta a_\mu^\Phil$ is proportional to $\lambda_{\Phi_{\ell}}^2$ and depends on the parameters $(m_M, \sin \delta_L, m_{\Phi_{\ell}})$. Taking $\lambda_{\Phi_{\ell}}=1$ for simplicity, we show in figure~\ref{fig:phil} the resulting $\Delta a^{\Phi_{\ell}}_{\mu}$ as a function of $m_{\Phi_{\ell}}$ for various values of $m_M$ and $\sin \delta_L$. It can be seen that the $\Phi_{\ell}$ contribution is always negative, which increases the discrepancy of $(g-2)_\mu$. However, such a negative contribution will be highly suppressed by heavy $\Phi_{\ell}$ or small $\lambda_\Phil$. In addition, by comparing the left and right plots in figure~\ref{fig:phil}, one finds that $|\Delta a _\mu^\Phil|$ becomes smaller for smaller $\sin\delta_L$. Especially, in the case of $m_\Phil>2\TeV$, $\lambda_\Phil<0.1$ and $\sin\delta_L<0.01$, we obtain $|\Delta a_\mu^\Phil|< 7.5\times 10^{-13}$, which can be safely neglected. In the following, we will consider such a special case.  

\item The $\phi$ contribution $\Delta a_\mu^\phi$ depends on the parameters $(\sin \delta_L, m_M, m_{\phi}, \lambda_\phi)$. We take $m_M=300\GeV$ as discussed in section~\ref{sec:Zmumu}, and show in figure~\ref{fig:g-2lamphi-mphi} the $2\,\sigma$ allowed regions of $(m_{\phi}, \lambda_{\phi})$ for $\sin \delta_L=(1.0,\, 0.5,\, 0.1)\times 10^{-2}$. It can be seen that the $\phi$ boson can explain the $(g-2)_\mu$ anomaly at the $2\,\sigma$ level. Furthermore, larger $\lambda_\phi$ is required for larger $m_\phi$ and the lower limits on $\lambda_\phi$ depend only marginally on $\sin\delta_L$ for large $m_\phi$. Taking $\lambda_\phi=2$, we can then derive the $2\,\sigma$ allowed region of $(m_\phi, \sin \delta_L)$, which is shown in the left panel of figure~\ref{fig:g-2ssl-mphi}. As can be seen from figures~\ref{fig:g-2lamphi-mphi} and \ref{fig:g-2ssl-mphi}, smaller $m_\phi$ corresponds to a larger range of $\sin\delta_L$ and $\lambda_\phi$. Hence, we take $m_\phi=1\GeV$ as a benchmark value and derive the allowed region of $(\lambda_\phi, \sin\delta_L)$, which is shown in the right panel of figure~\ref{fig:g-2ssl-mphi}.

\item The $\Zp$ contribution $\Delta a_\mu^\Zp$ depends on the parameters $(m_\Zp, m_M, \gl, \sin \delta_L)$ and is positive. However, different from the $\phi$-boson case, the $\Zp$ can also affect the neutrino trident production. After considering the constraint from this rare process, the $\Zp$ contribution alone is not sufficient to explain the $(g-2)_\mu$ anomaly, as observed in the minimal $U(1)_{L_\mu-L_\tau}$ model~\cite{Altmannshofer:2014pba}.

\end{itemize}

As discussed above, in the case of $m_\Phil>2\TeV$ and $\lambda_\Phil<0.1$, the relevant contributions to the $(g-2)_\mu$ arise from both the $\Zp$ and the $\phi$ boson. Thus, we consider them together, and take $m_M=300\GeV$, $m_\Zp=1000\GeV$ and $\gl=2$ for simplicity. After taking into account the constraints from the $(g-2)_\mu$ anomaly and the neutrino trident production, we can derive the $2\,\sigma$ allowed regions of $(\sin\delta_L,\, m_\phi,\, \lambda_\phi)$, which are also shown in figures~\ref{fig:g-2lamphi-mphi} and \ref{fig:g-2ssl-mphi}. One can see that the allowed regions become much larger compared to that obtained by including only the $\phi$ contribution. We take $m_\phi=1\GeV$ as a benchmark value, which approximately provides the largest parameter space for $\sin\delta_L$ and $\lambda_\phi$, and show in figure~\ref{fig:g-2ssl-mphi} the allowed region of $(\lambda_\phi,\, \sin\delta_L)$. In this case, the lower bound on $\sin\delta_L$ is $3.2\times 10^{-4}$ and, after combining the bound derived in section~\ref{sec:Zmumu}, we obtain
\begin{align}\label{eq:bound:slL}
 3.2 \times 10^{-4} <\sin \delta_L < 1.0 \times 10^{-2}\,.
\end{align}

It is noted that the $\Zp$ and the $\phi$ contribution are both positive, and the latter alone can account for the $(g-2)_\mu$ anomaly while providing no influence on other processes discussed in section~\ref{sec:theory}. Therefore, we will take $m_\phi=1\GeV$ and the bound in eq.~(\ref{eq:bound:slL}) in the following analysis. This in turn implies that we do not need to consider the $(g-2)_\mu$ anomaly anymore when investigating the $\Zp$ contributions to other processes, as the anomaly can be definitely resolved in our model.

\subsection{$b \to s \ell^+ \ell^-$ processes}

\begin{figure}[t]
  \centering
  \includegraphics[width=1.00\textwidth]{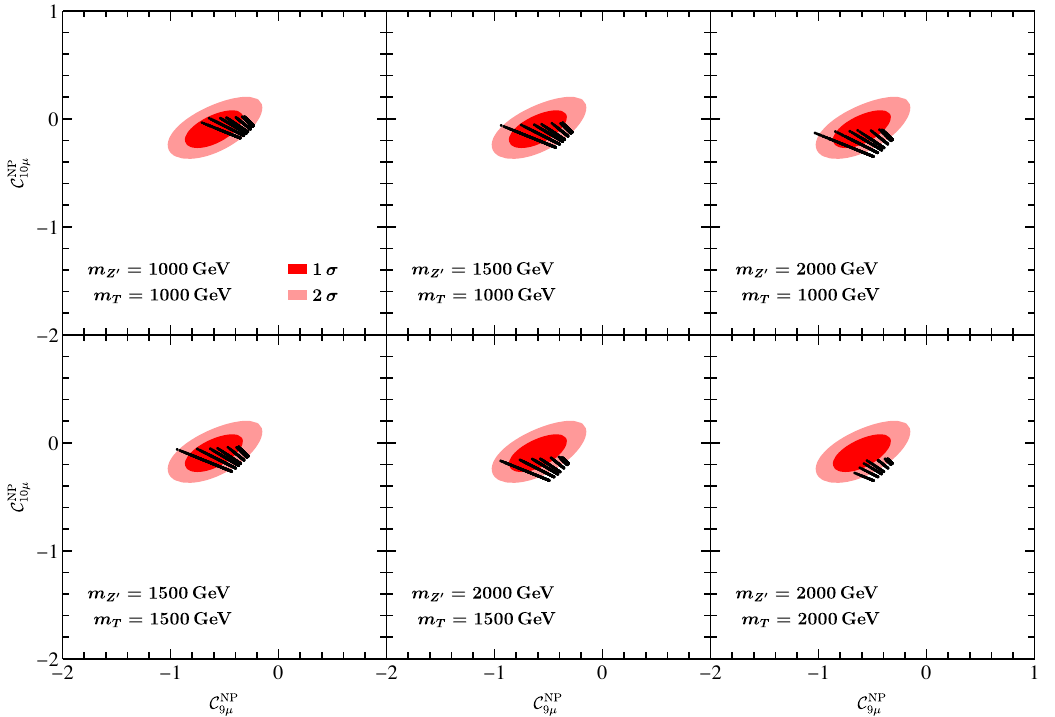}
  \caption{Constraints on $\big(\mC_{9\mu}^\NP, \mC_{10\mu}^\NP \big)$ from a global fit of the $b \to s \ell^+ \ell^-$ processes. The allowed regions at the $68\%$ and the $95\%$ confidence level (CL) are shown by the dark and the light red, respectively. The black points show the predicted values of $\mC_{9\mu}^\NP$ and $\mC_{10\mu}^\NP$ based on the whole allowed parameter space shown in figure~\ref{fig:PS:global:1}.}
  \label{fig:PS:C9C10}
\end{figure}

The most relevant observables to our model include the $b \to s \ell^+ \ell^-$ transitions,\footnote{The top partner can also affect the radiative $b \to s \gamma$ decays. However, the NP contribution is proportional to $\sin^2\theta_L$ and thus highly suppressed for small $\theta_L$.} the $W$-boson mass, the $(g-2)_{\mu}$, the neutrino trident production, as well as the $Z\mu\mu$ coupling. As discussed in the last subsection, the $\phi$ contribution alone can explain the $(g-2)_{\mu}$ anomaly and, at the same time, does not bring any significant effect on other observables. Therefore, the remaining question is to see if the parameter space required to account for the latest CDF $m_W$ measurement can also explain the $b \to s \ell^+ \ell^-$ discrepancies, while satisfying the constraints from the neutrino trident production and the $Z\mu\mu$ coupling. This will be explored in this subsection.

For the $b \to s \ell^+ \ell^-$ transitions, only the short-distance Wilson coefficients $\mC_9$ and $\mC_{10}$ are affected in our model. Therefore, we will perform a global fit of $\mC_{9\mu}^\NP$ and $\mC_{10\mu}^\NP$ by considering the various measurements of the $b \to s \ell^+ \ell^-$ processes, including the recent measurements of $R_{K^{(*)}}$~\cite{LHCb:2022qnv,LHCb:2022zom} and $\mB(B_s \to \mu^+ \mu^-)$~\cite{CMS:2022mgd}. Details of the fit are given in appendix~\ref{app:fit}. The final allowed regions of $(\mC_{9\mu}^\NP,\, \mC_{10\mu}^\NP)$ obtained through such a global fit are shown in figure~\ref{fig:PS:C9C10}, which are very similar to that of the latest global fits by other groups~\cite{Gubernari:2022hxn,Greljo:2022jac,Ciuchini:2022wbq,Alguero:2023jeh,Wen:2023pfq}.

\begin{figure}[t]
  \centering
  \includegraphics[width=1.00\textwidth]{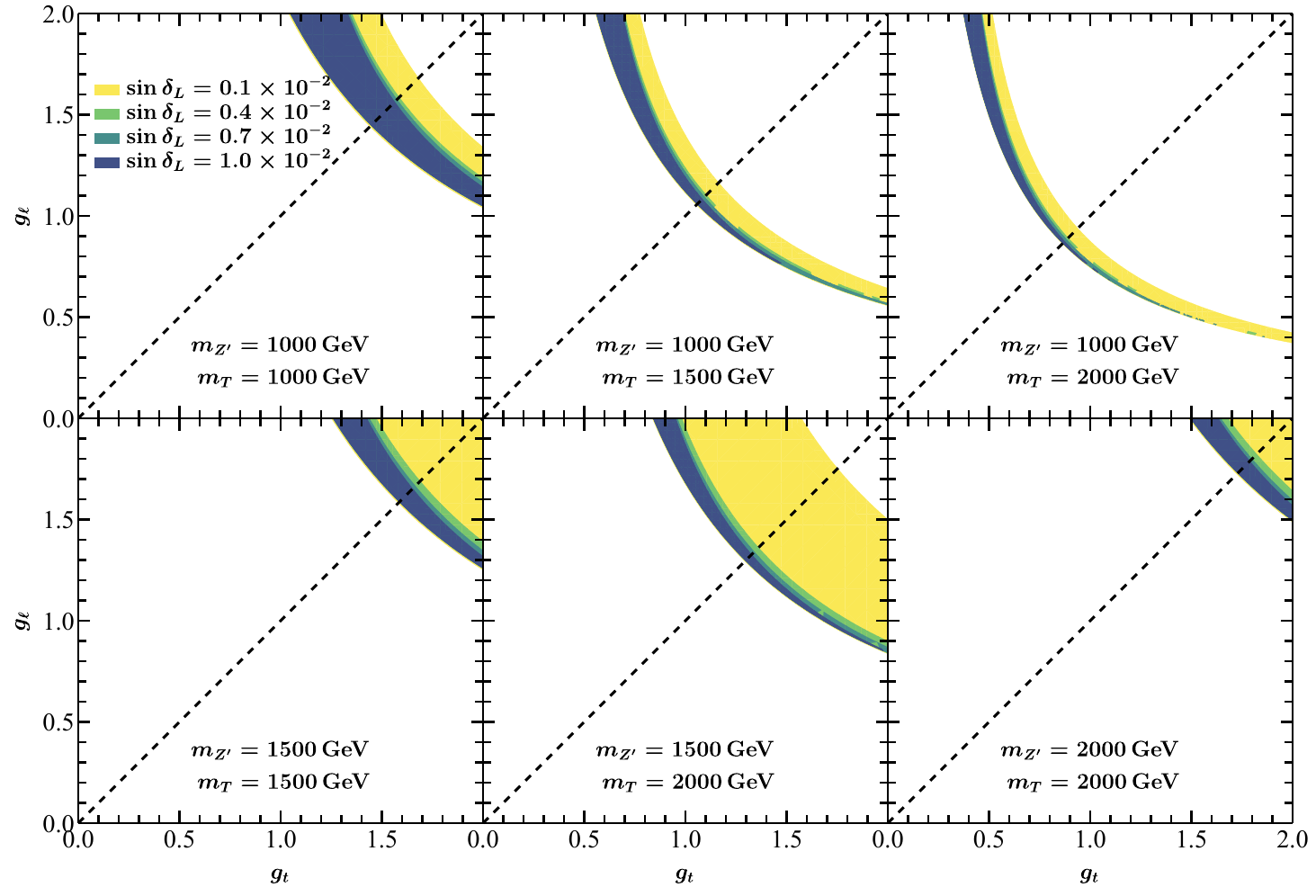}
  \caption{Constraints on $(\sin\theta_L,\, \sin\delta_L,\, \gt,\, \gl)$ for several benchmark values of $(m_\Zp,\, m_T)$, plotted in the $(\gt, \gl)$ plane. The $2\,\sigma$ allowed regions for $\sin\delta_L=(0.1,\, 0.4,\, 0.7,\, 1.0)\times 10^{-2}$ are represented by different colors.}
  \label{fig:PS:global:1}
\end{figure}

Since a small mixing angle $\theta_L$ is required to explain the latest CDF $m_W$ measurement, the NP contributions to $\mC_9$ and $\mC_{10}$ from the $W$-box, $\gamma$- and $Z$-penguin diagrams, which are all proportional to $\sin^2 \theta_L$, are highly suppressed. Therefore, the dominant contributions arise from the $\Zp$-penguin diagrams, which depend on the parameters $(m_T, \sin\theta_L, m_M, \sin\delta_L, m_\Zp, \gt, \gl)$. In the following, we take $m_M=300\GeV$ as discussed in section~\ref{sec:Zmumu}, and choose $(m_\Zp,\, m_T)=(1.0,\, 1.0)$, $(1.0,\, 1.5)$, $(1.0,\, 2.0)$, $(1.5,\, 1.5)$, $(1.5,\, 2.0)$, $(2.0,\, 2.0)\TeV$ as several benchmark values. Then, the remaining relevant parameters are $(\sin\theta_L,\sin\delta_L, \gt, \gl)$. Considering the $2\,\sigma$ allowed regions obtained from the global EW fit as well as the constraints from the $b \to s \ell^+ \ell^-$ global fit, the neutrino trident production and the $Z\mu\mu$ coupling,\footnote{For the constraint from the $Z\mu\mu$ coupling, we take $\lambda_\phi=\lambda_\Phil=1$, $m_\phi=1\GeV$ and $m_\Phil=2\TeV$, as discussed in the section~\ref{sec:Zmumu}.} we can finally derive the allowed regions of these parameters. As an illustration, we show in figure~\ref{fig:PS:global:1} the allowed regions in the $(\gt,\, \gl)$ plane for $\sin\delta_L=(0.1,\, 0.4,\, 0.7,\, 1.0)\times 10^{-2}$. It can be seen that the allowed values of the parameters $\gt$ and $\gl$ for TeV-scale $\Zp$ and $T$ are both of $\mathcal O(1)$ simultaneously, therefore being safely in the perturbative region. For heavier $T$ or lighter $\Zp$, smaller $\gt$ and $\gl$ are required. In addition, for larger $\sin\delta_L$, the allowed regions of $\gt$ and $\gl$ are reduced.

\begin{figure}[t]
  \centering
  \includegraphics[width=1.00\textwidth]{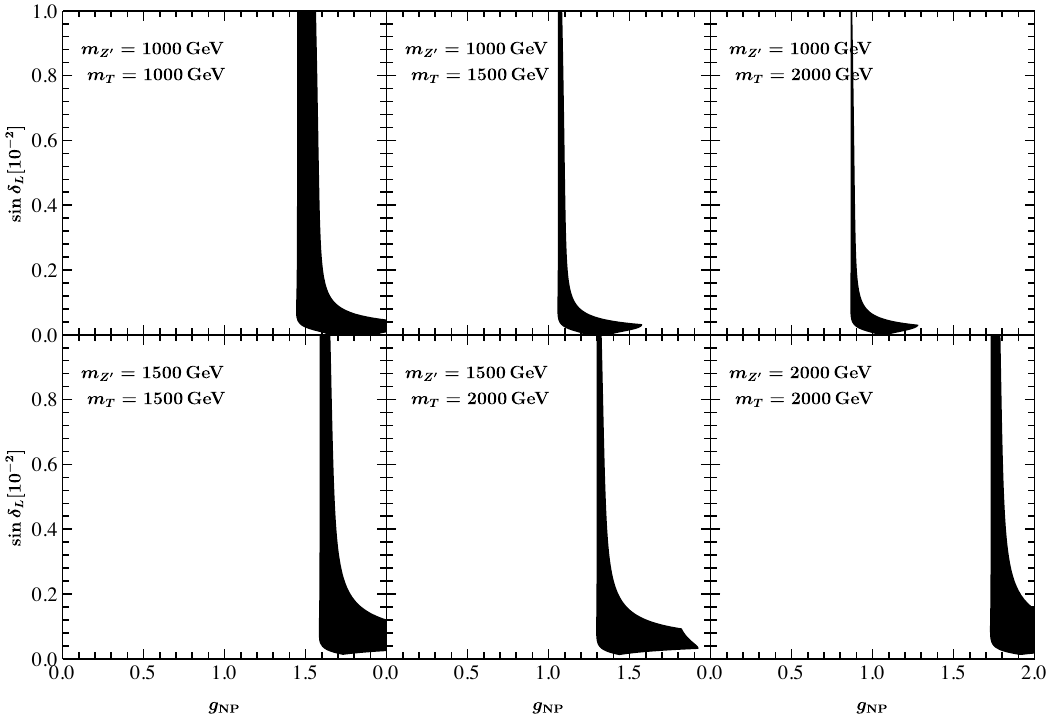}
  \caption{Constraints on the model parameters $(\sin\theta_L,\, \sin\delta_L,\, \gt,\, \gl)$ for several benchmark values of $(m_\Zp,\, m_T)$ in the case of $\gt=\gl\equiv g_\NP$, plotted in the $(g_\NP, \sin\delta_L)$ plane. The $2\,\sigma$ allowed regions are shown in black.}
  \label{fig:PS:global:2}
\end{figure}

As a conclusion, our model can accommodate the $(g-2)_{\mu}$ anomaly, the latest CDF $W$-mass shift and the $b \to s \ell^+ \ell^-$ discrepancies, while satisfying the other constraints like the neutrino trident production and the $Z\to\mu^+\mu^-$ processes. With the model parameters varied within the allowed regions, the predicted Wilson coefficients $\mC_{9\mu}^\NP$ and $\mC_{10\mu}^\NP$ are also shown in figure~\ref{fig:PS:C9C10}. From figure~\ref{fig:PS:global:1}, we can also see that the departure of the ratio $\gl/\gt$ from unity is more stringently constrained for heavier $\Zp$ or lighter $T$. Furthermore, for different values of $(m_\Zp,\, m_T)$, $\gt = \gl$ (or equivalently $q_\ell = q_t$) is always allowed. In the case of $\gt=\gl\equiv g_\NP$, we show in figure~\ref{fig:PS:global:2} the allowed parameter space in the $(g_\NP,\, \sin\delta_L)$ plane. Notably, this in turn implies that the possibility of $\Phit = \Phi_\ell^*$ is allowed in our model.\footnote{The possibility of $\Phit=\Phil$ is, however, not possible, since $q_\ell=-q_t$ makes the sign of $\mC_{9\mu}$ and $\mC_{10\mu}$ (cf. eq.~\eqref{eq:WC:Zp}) opposite to that shown in figure~\ref{fig:PS:C9C10}.}

\subsection{Collider phenomenology}
\label{collider}

\begin{figure}[t]
  \centering
  \includegraphics[width=0.29\textwidth]{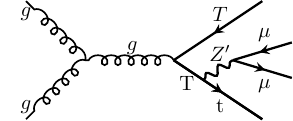}
  \qquad
  \includegraphics[width=0.29\textwidth]{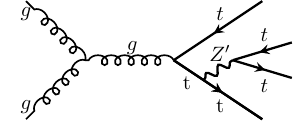}
  \qquad
  \includegraphics[width=0.29\textwidth]{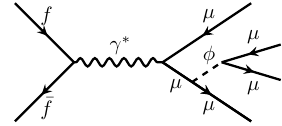}
  \caption{Representative Feynman diagrams for the processes $p p \to \mu ^+ \mu^- + X$ (left), $p p \to t \bar t t \bar t$ (middle), and $p p\,(e^+ e^-)\to \mu^+ \mu^- \mu^+ \mu^-$ (right) induced by the NP particles in our model.}
  \label{fig:collider:Feynman diagram}
\end{figure}

\begin{figure}[t]
	\centering
	\includegraphics[width=0.46\textwidth]{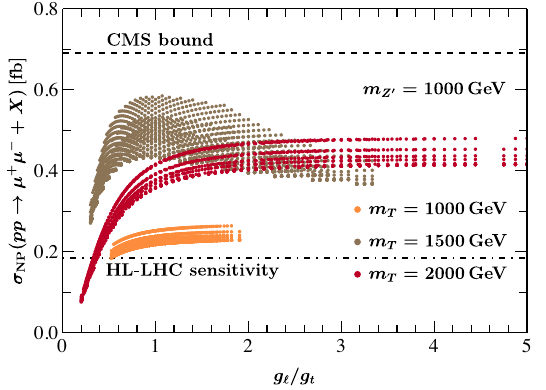}
	\qquad
	\includegraphics[width=0.46\textwidth]{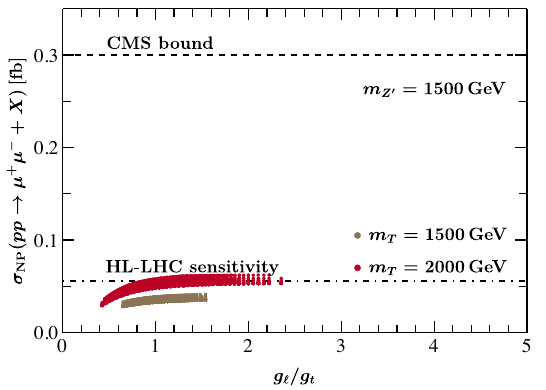}
	\\[0.2cm]	
	\includegraphics[width=0.46\textwidth]{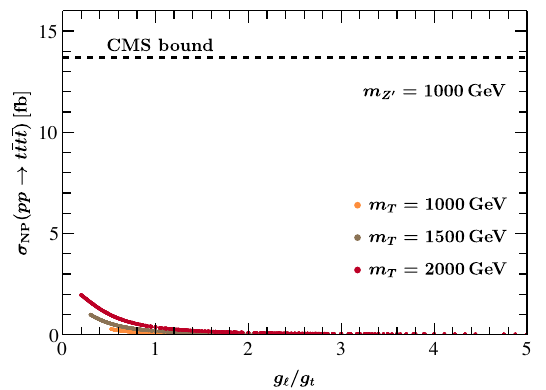}
	\qquad
	\includegraphics[width=0.46\textwidth]{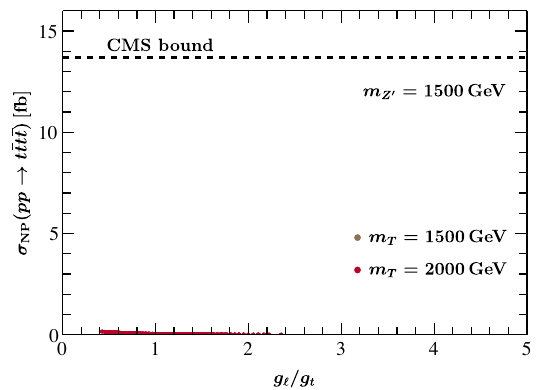}
	\caption{Predicted cross sections of the processes $pp \to \mu^+ \mu^-+X$ (top) and $pp \to t \bar t t \bar t$ (bottom), as well as the $95\%$ CL upper bounds from the CMS experiment~\cite{CMS:2023ftu,CMS:2021ctt}. For the $t \bar t t \bar t$ channel, the CMS upper bound is obtained by comparing the CMS measurement $17.7_{-3.5}^{+3.7}\text{(stat)}_{-1.9}^{+2.3}\text{(syst)}\fb$~\cite{CMS:2023ftu} with the SM prediction $13.4_{-1.8}^{+1.0}\fb$~\cite{vanBeekveld:2022hty}. For the dimuon channel, the sensitivities expected at the HL-LHC corresponding to $3000\fb^{-1}$ at $\sqrt s =14 \TeV$~\cite{CMS:2022gho} are also shown.}
	\label{fig:collider}
\end{figure}

The vector-like top partner, being colored, can be efficiently produced at the hadron colliders~\cite{Aguilar-Saavedra:2013wba,Greiner:2014qna,Cox:2015afa,Kim:2016plm}. Searches for single and pair productions of the vector-like top partner have been performed at the LHC, and strong constraints on its mass and mixing angle have been obtained~\cite{ATLAS:2018ziw,CMS:2022yxp,delaTorreTrishaFarooque:2022vqc}. However, in most of these searches, it is assumed that the top partner decays exclusively into the SM particles, \textit{i.e.}, $T \to bW/tZ/th$, and the top partner has been excluded for masses below $1.3\TeV$. In our model, this lower bound applies only for the case of $m_T < m_\Zp + m_t$. In the case of $m_T > m_\Zp + m_t$, on the other hand, the $T \to t \Zp$ channel is open, and the bounds from these direct searches could be therefore relaxed~\cite{Serra:2015xfa,Anandakrishnan:2015yfa,Bizot:2018tds}. In this case, as shown in figure~\ref{fig:collider:Feynman diagram}, the cascade decay $T \to t\Zp (\to \mu^+ \mu^-)$ makes the dimuon resonance searches at the LHC~\cite{ATLAS:2019erb,CMS:2021ctt} sensitive to our model. In order to derive the collider constraints, we take $m_\phi = 1\GeV$, $m_\Phit=m_\Phil=2\TeV$ and $m_M = 300\GeV$, and consider the allowed parameter space of $(\sin\theta_L,\, \sin\delta_L,\, g_t,\, g_\ell)$ derived in the last subsection, which corresponds to that shown in figure~\ref{fig:PS:global:1}. For the dimuon resonance searches, the cross section is estimated by $\sigma(p p \to T \bar T)\cdot 2 \cdot \mB(T \to t \Zp)\cdot \mB(\Zp \to \mu^+ \mu^-)$ and the result of $\sigma(p p \to T \bar T)$ in ref.~\cite{ATLAS:2018ziw} is used. The total width of the top partner is calculated by considering all the tree-level two-body decay modes, which consist of $T\to t Z^{(\prime)}$, $b W$ and $th$. Here the main decay channel is $T \to t \Zp $ for small $q_\ell/q_t$ and $T \to t h$ for large $q_\ell/q_t$. For the $\Zp$ boson, its total width can be estimated by including all the tree-level two-body decay modes, which consist of $\Zp \to M^+ M^-$, $M^+ \mu^-$, $\mu^+ M^-$, $\mu^+ \mu^-$, $\tau^+ \tau^-$, $\nu_\mu \bar\nu_\mu$, $\nu_\tau \bar\nu_\tau$ and $t \bar t$. The main decay channel is $\Zp \to t \bar t$ for small $q_\ell /q _t$. For large $q_\ell /q_t$, on the other hand, the $\Zp$ decay is dominated by $\Zp \to \tau^+ \tau^-$ and $\Zp \to \nu_{\mu,\tau}\bar\nu_{\mu,\tau}$, while the branching ratio $\mB(\Zp \to \mu^+ \mu^-)$ can at most reach about $20\%$. The cross sections corresponding to the allowed parameter space can then be derived, which are shown as a function of $q_\ell/q_t$ in figure~\ref{fig:collider}. It can be seen that the cross sections for various values of $m_\Zp$ and $m_T$ are all below the current CMS bound~\cite{CMS:2021ctt}. The cross section becomes larger for a lighter $\Zp$; especially for $m_T=1.0\TeV$, the maximum cross section is close to the CMS bound. Furthermore, in most of the parameter space, the cross sections are higher than the sensitivities expected at the High-Luminosity LHC (HL-LHC) corresponding to $3000\fb^{-1}$ at $\sqrt s = 14\TeV$~\cite{CMS:2022gho}.

Searches for the multi-top final states~\cite{CMS:2023ftu,ATLAS:2023ajo} can also provide evidence of the $\Zp$ boson~\cite{Fox:2018ldq}. As shown in figure~\ref{fig:collider:Feynman diagram}, the $\Zp$ boson in this case can be produced in association with top pairs, and the decay $\Zp \to t \bar t$ leads to the four-top final state $t \bar t t \bar t$. Similar to the analysis of the dimuon channel, we consider the allowed parameter space corresponding to figure~\ref{fig:PS:global:1}. Predictions of the NP contributions to the $t \bar t t \bar t$ cross section are shown in figure~\ref{fig:collider}, which are estimated by $\sigma_\NP(p p \to t \bar t t \bar t) \approx \sigma(pp \to t \bar t \Zp)\cdot \mB(\Zp \to t \bar t)$, with the result of $\sigma(p p \to t \bar t \Zp)$ taken from ref.~\cite{Fox:2018ldq}. It can be seen that our predictions for various values of $m_\Zp$ and $m_T$ are all well below the current CMS bound.\footnote{The recent ATLAS measurement~\cite{ATLAS:2023ajo} is roughly $1.8\,\sigma$ higher than the SM prediction~\cite{vanBeekveld:2022hty}. For small $\gl/\gt$, $\sigma(p p \to t \bar t t \bar t)$ can be enhanced by $10\sim15\,\%$ compared to that in the SM. Thus, the tension is relaxed in our model.} Furthermore, the dimuon channel is enhanced by the NP contribution for large $q_\ell/q_t$, while the four-top channel enhanced by small $q_\ell/q_t$. Therefore, the two processes are complementary to each other in searching for the $\Zp$ boson.

Similar to the top partner, the vector-like lepton partner can also be pair-produced at the LHC. However, the production occurs only through the $s$ channel involving the EW vector bosons, leading to much smaller cross sections~\cite{Kumar:2015tna,Bhattiprolu:2019vdu}. Searches for vector-like leptons have been performed at the LEP~\cite{L3:2001xsz} and the LHC experiment~\cite{CMS:2022nty,ATLAS:2023sbu}, and most of these studies focus on the case of vector-like tau partner. For example, by using $138\fb^{-1}$ of $pp$ collisions at $\sqrt{s}=13\TeV$, the $SU(2)_L$ doublet and singlet vector-like tau partner are already excluded for masses below $1045\GeV$ and in the mass range of $125-150\GeV$, respectively~\cite{CMS:2022nty}. In our model, due to the small mixing angle $\delta_L$, the particle $M$ can be approximated as a singlet vector-like muon partner. Similar to the case of the singlet vector-like tau partner, searches for the singlet vector-like muon partner is also very challenging for the LHC, due to its small production cross section. Based on the ATLAS searches for the anomalous productions of multi-lepton events~\cite{ATLAS:2013swe}, the vector-like muon partner has been investigated in ref.~\cite{Dermisek:2014qca} (see ref.~\cite{Falkowski:2013jya} for similar study by using the CMS data~\cite{CMS:2013jfa}). However, no limits on the vector-like muon partner are found for masses below $105\GeV$ or above $300\GeV$. In addition, the LEP limit on additional heavy leptons~\cite{L3:2001xsz} places a lower bound of around $100\GeV$ on the mass of the vector-like muon partner. Confronted with these experimental status, the benchmark value $m_M=300\GeV$ used in our numerical analysis is therefore reasonable. In addition, the muon partner lighter than around $300\GeV$ could produce a signal at the future proton-proton colliders~\cite{Bhattiprolu:2019vdu}. Detailed analysis of the prospects for its discovery will be presented in a future work.

Besides the top partner, the muon partner and the $\Zp$ boson, signals of the scalars $\Phit$, $\Phil$ and $\phi$ can also be searched for at the high-energy colliders. For example, as shown in figure~\ref{fig:collider:Feynman diagram}, the scalar $\phi$ can be produced in association with a muon pair, and the decay $\phi \to \mu^+ \mu^-$ leads to a four-muon final state, which has been measured at the LHC~\cite{CMS:2018yxg,ATLAS:2023vxg} and the $B$ factories~\cite{BaBar:2016sci,Belle:2021feg}. For $m_\phi=1\GeV$, which has been used in our numerical analysis, future measurements at the Belle II~\cite{Laurenza:2022rjm} and the Super Tau-Charm Facility (STCF)~\cite{Achasov:2023gey} are expected to provide sensitive probes of this scalar. Detailed analysis of the current constraints on the scalars in our model and the future prospects at the LHC, Belle II and STCF is, however, beyond the scope of this paper, but will be explored in our future work.

\section{Conclusion}
\label{sec:conclusion}

In this paper, we have constructed a NP model that successfully addresses the latest CDF $W$-boson mass shift, the $(g-2)_\mu$ anomaly and the $b \to s \ell^+ \ell^-$ discrepancies. In our setup, the SM is extended by the $SU(2)_L$-singlet vector-like top and muon partners that are featured by additional $U(1)^\prime$ gauge symmetry. The top and the muon partner have also the same SM quantum numbers as of the right-handed top and muon respectively, and can therefore mix with the latter after the EW and the $U(1)^\prime$ symmetry breaking.

Similar to the SM case, the loop diagrams involving these fermion partners can contribute to the $(g-2)_\mu$, the $W$-boson mass $m_W$ and the $b \to s \mu^+ \mu^-$ transitions. After considering the most relevant constraints, such as the $Z \to \mu^+ \mu^-$ decay and the neutrino trident production, both the $(g-2)_\mu$ anomaly and the latest CDF $W$-boson mass shift can be explained in our model. This requires that the mixing angles between the fermions and the fermion partners should be small. At the same time, the $Z^\prime$-penguin diagrams involving the top partner can affect the short-distance Wilson coefficients $\mC_{9\mu}$ and $\mC_{10\mu}$ in the $b \to s \mu^+ \mu^-$ transitions. Furthermore, the small lepton mixing angle $\delta_L$ makes the $\Zp \mu \mu$ interaction almost of a vector-type. Therefore, $\mC_{9\mu}^\NP \ll \mC_{10\mu}^\NP\approx 0$ is obtained in our setup. This is also favored by the $b \to s \ell^+ \ell^-$ global fit after including the recent measurements of $R_{K^{(*)}}$~\cite{LHCb:2022qnv,LHCb:2022zom} and $\mB(B_s \to \mu^+ \mu^-)$~\cite{CMS:2022mgd}. It is also found that both the $\Zp$ boson and the top partner can be as light as of $1 \sim 2\TeV$, and thus may be accessible at the LHC Run 3 and its upgrade. Searches for the dimuon resonances and the top partner could also provide evidences of the $\Zp$ boson and the top partner. Especially, the $\Zp$ boson can enhance the $p p \to t \bar t t \bar t$ production cross section. Finally, the scalar $\phi$ with a mass of around $1\GeV$ can be produced in association with a muon pair, and the subsequent decay $\phi \to \mu^+ \mu^-$ leads to a four-muon final state, which can be searched for at the Belle II and STCF experiments.

As a final comment, our model can be further explored in several phenomenological directions. By allowing for nonzero mixings of the top (muon) partner with the first and second (third) generations of quarks (leptons), several other interesting observables could be affected. In particular, the CKM matrix should be extended in this case, which may be responsible for the Cabibbo angle anomaly~\cite{Belfatto:2021jhf,Crivellin:2022rhw}. Our model can also be extended by right-handed neutrinos or a dark sector. These points will be investigated in our future works.

\acknowledgments

This work is supported by the National Natural Science Foundation of China under Grant Nos. 12135006, 12075097 and 11805077, as well as by the Fundamental Research Funds for the Central Universities under Grant Nos. CCNU19TD012, CCNU20TS007 and CCNU22LJ004. XY is also supported in part by the Startup Research Funding from CCNU.

\appendix

\section{\boldmath One-loop corrections to $Z \to \ell^+ \ell^-$}
\label{app:Zmumu}

At the one-loop level, the NP contributions to the $Z\to\mu^+\mu^-$ decay arise from the diagrams shown in figure~\ref{fig:diagram:Zmumu}. In our calculation, the on-shell renormalization scheme specified in ref.~\cite{Denner:1991kt} is adopted. Furthermore, in order to cancel the ultraviolet (UV) divergences, the renormalization of the mixing angle $\delta_{L}$ should be performed. In the following, details of the renormalizations of the lepton fields and the mixing angle are given, respectively.

In the mass eigenbasis, the fermion field renormalization is performed through
\begin{align}\label{ren}
f^{L/R}_{i,0}&=(Z^{L/R}_{ij})^{\frac{1}{2}} f^{L/R}_j\equiv(\delta_{ij}+\frac{1}{2}\delta Z^{L/R}_{ij})f^{L/R}_j\,,
\end{align} 
where $f^{L/R}_{i}$ and $f^{L/R}_{i,0}$ denote the renormalized and bare left/right-handed lepton fields, respectively. $Z^{L/R}_{ij}$ represent the fermion field renormalization matrices, and can be fixed by the following on-shell renormalization conditions:
\begin{align}
  \hat{\Gamma}_{i j}(p) u_j(p)\Big |_{p^2=m_{j}^2}=0,
  \quad\quad
 \lim _{p^2 \to m_{i}^2} \frac{\slashed p+m_{i}}{p^2-m_{i}^2}  \hat{\Gamma}_{i i}(p) u_i(p)=i u_i(p),
\end{align}
where $u_i(p)$ denotes the spinor of the external fermion fields. The renormalized one-particle irreducible two-point function $\hat{\Gamma}_{i j}$ is defined by
\begin{align}
    \hat{\Gamma}_{i j}(p) &= \parbox[c]{2cm}{\includegraphics[width=3cm]{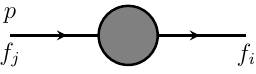}}\qquad \quad, \nonumber \\
	& = i \delta_{i j}\left(\slashed{p}-m_i\right) + i \left[\slashed{p} P_L \Sigma_{i j}^L(p^2)+\slashed{p} P_R \Sigma_{i j}^R(p^2)+P_L \Sigma_{i j}^{SL}(p^2)+P_R \Sigma_{i j}^{SR}(p^2)\right]\,,
\end{align}
where the scalar functions $\Sigma_{i j}^{L}$, $\Sigma_{i j}^{R}$, $\Sigma_{i j}^{SL}$ and $\Sigma_{i j}^{SR}$ are functions of $p^2$. Then, the on-shell renormalization conditions yield
\begin{align}\label{fercon}
	\delta Z_{i j}^{L}=&\frac{2}{m_{i}^2-m_{j}^2} \operatorname{Re}\left[m_{j}^2 \Sigma_{i j}^{L}\left(m_{j}^2\right)+m_{i} m_{j} \Sigma_{i j}^{R}\left(m_{j}^2\right)+m_{i} \Sigma_{i j}^{SL}\left(m_{j}^2\right)+m_{j} \Sigma_{i j}^{SR}\left(m_{j}^2\right)\right], \quad \mbox{for $i \neq j$}, \\
	\delta Z_{i j}^{R}=&\frac{2}{m_{i}^2-m_{j}^2} \operatorname{Re}\left[m_{j}^2 \Sigma_{i j}^{R}\left(m_{j}^2\right)+m_{i} m_{j} \Sigma_{i j}^{L}\left(m_{j}^2\right)+m_{j} \Sigma_{i j}^{SL}\left(m_{j}^2\right)+m_{i} \Sigma_{i j}^{SR}\left(m_{j}^2\right)\right], \quad \mbox{for $i \neq j$},  \\
	\delta Z_{i i}^{L/R}=&-\operatorname{Re} \Sigma_{i i}^{L/R}\left(m_{i}^2\right)-\left.m_{i}^2 \frac{\partial}{\partial p^2} \operatorname{Re}\left[\Sigma_{i i}^{L/R}\left(p^2\right)+\Sigma_{i i}^{R/L}\left(p^2\right)+\frac{2}{m_{i}} \Sigma_{i i}^{SL/SR}\left(p^2\right)\right]\right|_{p^2=m_{i}^2}.
\end{align}
Specific to our model, their explicit expressions are also given in the ancillary notebook file.

After performing the field renormalization, the one-loop contribution to $g_{R\mu}$ in eq.~(\ref{eq:zmumulr}) is already finite. However, the one-loop contribution to $g_{L\mu}$ is still divergent. Here the renormalization of the mixing angle $\delta_{L}$ is required to cancel the remaining divergence. This is similar to the case in the Two-Higgs-Doublet Model, where the mixing angle relating the light and heavy neutral scalars should also be renormalized~\cite{Denner:2018opp,Kanemura:2004mg}. The renormalization of the mixing angle $\delta_L$ reads
\begin{align}
\delta_{L,0}=\delta_L+\Delta \delta_L,
\end{align}
where $\delta_{L,0}$ and $\delta_L$ denote the bare and renormalized mixing angles, respectively. Then, in terms of the renormalized quantities, the $Z$ and $W$ interactions with the left-handed leptons in eqs.~(\ref{eq:interaction:l:W}) and (\ref{eq:interaction:l:Z}) become
\begin{align}
\Delta \mathcal L_W^\ell =&\frac{g}{\sqrt{2}} \left[\left(-\hat{s}_L \Delta \delta_L+\frac{\hat{c}_L}{2} \delta Z_{\mu\mu}^L+\frac{\hat{s}_L}{2} \delta Z_{M\mu}^L\right) \bar{\mu} \slashed{W} P_{L} \nu_{\mu}\right.\nonumber\\
&\quad \left. +\left(\hat{c}_L \Delta \delta_L+\frac{\hat{s}_L}{2} \delta Z_{MM}^L+\frac{\hat{c}_L}{2} \delta Z_{\mu M}\right) \bar{M} \slashed{W} P_{L} \nu_{\mu}\right]+\text { h.c.}\,,\\
\Delta \mathcal L_Z^\ell =&\frac{g}{2c_W}\left(\bar{\mu}, \bar{M}\right) \left[ \left (
                               \begin{pmatrix}
 2 \hat s_L \hat c_L & \hat s_L^2 - \hat c_L^2 \\ 
 \hat s_L^2 - \hat c_L^2 & - 2 \hat s_L \hat c_L
\end{pmatrix}
                           \Delta\delta_L
          +\hat C \hat Z_L + \hat Z_L^T \hat C \right )
          \slashed{Z} P_L
          + s_W^2 \big( \hat Z_R + \hat Z_R^T \big)
          \slashed{Z} P_R
          \right ]
          \begin{pmatrix}
\mu \\
M
\end{pmatrix},
\end{align}
with
\begin{align}
  \hat C =
  \begin{pmatrix}
-\frac{1}{2} \hat{c}_L^2+ s_W^2 & -\frac{1}{2} \hat{s}_L\hat{c}_L \\
-\frac{1}{2} \hat{s}_L\hat{c}_L & -\frac{1}{2} \hat{s}_L^2+s_W^2
\end{pmatrix},
  \quad\quad
          \hat Z_{L/R} =
          \begin{pmatrix}
 Z_{\mu\mu}^{L/R} & Z_{\mu M}^{L/R} \\ 
 Z_{M \mu}^{L/R} & Z_{MM}^{L/R}
\end{pmatrix}.
\end{align}
In order to fix the renormalization of the mixing angle $\delta_L$, we require the one-loop renormalized matrix element for the $Z\to\mu^+\mu^-$ decay is finite, \textit{i.e.},
\begin{align}\label{zp}
\mathcal{M}(Z \to \mu^+ \mu^-)|_\mathrm{UV}=0,
\end{align}
By calculating the vertex-correction diagrams shown in figure~\ref{fig:diagram:Zmumu}, we find that this renormalization condition leads to the following renormalization constant:
\begin{align}
\Delta \delta_L=-\frac{1}{2}\frac{\hat{c}_L}{\hat{s}_L}\delta Z_{\mu \mu}^L \big |_\mathrm{UV}-\frac{1}{2} \delta Z_{\mu M}^L \big |_\mathrm{UV}.
\end{align}
It is also found that, with these prescriptions, the one-loop renormalized matrix elements for the decays $Z\to\ell^+_i\ell^-_j$ and $W^+\to\ell^+_i\nu$~($\ell_{i,j}=e$ or $\mu$) are all finite after including the above renormalization constants.

\section{\boldmath Global fit of $b \to s \ell^+ \ell^-$ transitions}
\label{app:fit}

To perform a global fit of the $b \to s \ell^+ \ell^-$ transitions, as done in refs.~\cite{Altmannshofer:2017fio,Altmannshofer:2017yso,Aebischer:2019mlg,Altmannshofer:2021qrr}, we have considered the following experimental data: 1) the branching ratios of $B \to K \mu^+\mu^-$~\cite{LHCb:2014cxe}, $B \to K^*\mu^+\mu^-$~\cite{CDF:2012qwd,LHCb:2014cxe,CMS:2015bcy,LHCb:2016ykl}, $B_s \to \phi \mu^+\mu^-$~\cite{LHCb:2021zwz}, $\Lambda_b \to \Lambda \mu^+ \mu^-$~\cite{LHCb:2015tgy}, $B \to X_s \mu^+\mu^-$~\cite{BaBar:2013qry}, and $B_s \to \mu^+\mu^-$~\cite{LHCb:2021vsc,CMS:2022mgd}; 2) the angular distributions in $B \to K \mu^+\mu^-$~\cite{CDF:2012qwd,LHCb:2014auh}, $B \to K^*\mu^+\mu^-$~\cite{CMS:2015bcy,Belle:2016fev,CMS:2017ivg,ATLAS:2018gqc,LHCb:2020lmf,LHCb:2020gog}, $B_s \to \phi \mu^+\mu^-$~\cite{LHCb:2021xxq}, and $\Lambda_b \to \Lambda \mu^+ \mu^-$~\cite{LHCb:2018jna}; 3) the LFU ratios $R_{K}$ and $R_{K^{*}}$~\cite{BaBar:2012mrf,Abdesselam:2019wac,BELLE:2019xld,LHCb:2022qnv,LHCb:2022zom,LHCb:2021lvy}.

During the global fit, we firstly construct a likelihood function that depends only on the short-distance Wilson coefficients $\boldsymbol{\mC}$~\cite{Altmannshofer:2014rta}, 
\begin{align}
-2\ln L(\boldsymbol{\mC}) = \boldsymbol{x}^T (\boldsymbol{\mC}) \big[ V_{\rm exp} + V_{\rm th}(\boldsymbol{\theta})\big]^{-1} \boldsymbol{x}(\boldsymbol{\mC}),
\end{align}
where $\boldsymbol{x}_i(\boldsymbol{\mC})=\boldsymbol{\mO}_i^{\rm fexp} - \boldsymbol{\mO}_i^{\rm th}(\boldsymbol{\mC}, \boldsymbol{\theta})$, with $\boldsymbol{\mO}_i^{\rm th}$ and $\boldsymbol{\mO}_i^{\rm exp}$ representing the central values of the theoretical predictions and the experimental measurements, respectively. Their values depend on the input parameters $\boldsymbol{\theta}$ and the Wilson coefficients $\boldsymbol{\mC}$. Here $V_{\rm exp}$ and $V_{\rm th}$ denote the covariance matrices of the experimental measurements and the theoretical predictions, respectively. All the theoretical uncertainties and their correlations are included in $V_{\rm th}$. Approximately, $V_{\rm th}$ can be obtained by fixing the Wilson coefficients $\boldsymbol{\mC}$ to their default values within the SM. Furthermore, the theoretical uncertainties are approximated as Gaussian, which can be obtained by random samplings of the probability density functions of the input parameters $\boldsymbol{\theta}$. Finally, the $\Delta\chi^2$ function can be written as $\Delta\chi^2(\boldsymbol{\mC})=-2\ln L(\boldsymbol{\mC})/L_{\rm max}$, where $L_{\rm max}$ represents the maximum value of the likelihood function $L$ for different values of the Wilson coefficients $\boldsymbol{\mC}$. More details about the fitting procedures can be found in refs.~\cite{Altmannshofer:2014rta,Straub:2018kue,Li:2021qyo}. Here we have used an extended version of the package \texttt{flavio}~\cite{Straub:2018kue} when performing such a global fit.

\section{Loop functions}
\label{sec:loop}

Explicit expressions of the loop functions present in the oblique parameters $S, T, U$ in eqs.~\eqref{eq:S}--\eqref{eq:U} are listed below:
\begin{align}
K_1(x,y) = & -44x + 2\ln x + f_{-9}(x,x) - (x \to y), \\
K_2(x,y) =&  (x-y)^2 -3x -\left[(x-y)^3 - 3(x-y) - \frac{6xy}{x-y} \right]\ln x -f_3(x,x) + f_3(x,y) + (x \leftrightarrow y), \\
K_3(x,y) =& -3 x - 6 x^2 +\left(8-18x+6x^3\right)\ln x - \left(12 - 18 x+ 6x^3\right) \ln(x-1) -32 y + 4 f_0(y, y),\\
K_4(x,y) = & -20x + 2\ln x + f_{-3}(x,x) - (x \to y), \\
K_5(x,y) =& \frac{1}{x-y}\left[-3x^2+2 x^3-6 x^2 y \right]-\left[(x-y)^3+3(x+y)+\frac{6 x^2}{-x+y}\right] \ln x  \nonumber
\\&-f_3(x,x)+f_3(x,y)+ (x \leftrightarrow y), \\
K_6(x,y) =& -x - 2 x^2 +2\left(2-3x+x^3\right)\ln\left[\frac{x}{x-1}\right]  -16 y + 2 f_0(y, y),
\end{align}
with the function $f_n(x,y)$ defined by
\begin{align}
f_n(x,y) = -[(x+y-1)^2-4xy-3+n(x+y)] \sqrt{4xy-(x+y-1)^2}\cos^{-1}\bigg(\frac{x+y-1}{\sqrt{4xy}}\bigg).
\end{align}

Explicit expressions of the loop integrals $I_{1-7}$ present in the short-distance Wilson coefficients $\mC_9^{\rm NP}$ and $\mC_{10}^{\rm NP}$ in section~\ref{bsll} are given, respectively, by
\begin{align}
I_1 & = f_t\big (-\tfrac{17}{6},-\tfrac{7}{24},\tfrac{1}{4}\big ) +\big[-\tfrac{1}{4}+f_t\big (3,3,\tfrac{1}{6},-\tfrac{1}{4}\big)\big]\ln x_t - (t \to T), \\[0.2cm]
I_2 & = \tfrac{3}{4}x_t-f_t \big( \tfrac{9}{4} \big)+ \big[ \tfrac{9}{4}+f_t \big (\tfrac{9}{2},\tfrac{9}{4} \big ) \big]\ln x_t - (t \to T), \\[0.2cm]
I_3 & = 3(x_t+x_T)+\frac{3}{2}\frac{x_tx_T}{x_T-x_t}\ln\frac{x_t}{x_T}, \\[0.2cm]
I_4 & = \frac{3x_t}{4} \left[ \frac{1}{3}-f_t(1) +2g(x_t,x_T)\ln x_t + (t \leftrightarrow T) \right], \\[0.2cm]
I_5 & = \frac{x_t^2}{2}\left[\frac{1}{x_t} +\frac{2\ln x_t}{x_T-x_t} + (t \leftrightarrow T) \right], \\[0.2cm]
I_6 &=-\frac{(x_t-x_T)(x_t x_T-x_T-x_t+4)}{2(x_t-1)(x_T-1)}-3x_t^2f_t(1)\ln{x_t}-(t \leftrightarrow T),\\[0.2cm]
I_7 &=-(x_t+x_T)+2\frac{x_tx_T}{x_t-x_T}\ln\frac{x_t}{x_T}.
\end{align}
where $x_t=m_t^2 / m_W^2$, $x_T=m_T^2 / m_W^2$, and the functions $g(x,y)$ and $f_q(a_1, a_2, a_3, \dotsc, a_n)$ are defined, respectively, as
\begin{align}
  g(x,y)=\frac{y(4-8y+y^2)+x(4-2y+y^2)}{6(x-y)(y-1)^2},
  \qquad 
  f_q(a_1, a_2, a_3, \dotsc, a_n) = \sum_{i=1}^n\frac{a_i}{(x_q-1)^i}.
\end{align}

The loop functions $I_L$ and $I_R$ introduced in eq.~(\ref{eq:g-2:mixing}) are given, respectively, by
\begin{align}
I_L = - (m_t^2+m_T^2) + \frac{2m_t^2 m_T^2}{m_t^2-m_T^2}\ln\frac{m_t^2}{m_T^2},
\qquad
I_R = 4m_T^2 - \frac{2(m_t^2+m_T^2)m_T^2}{m_t^2-m_T^2}\ln\frac{m_t^2}{m_T^2}.
\end{align}

\bibliographystyle{JHEP}

\begingroup
\setlength{\bibsep}{5pt}
\bibliography{ref}
\endgroup

\end{document}